\documentclass[prd,preprint,superscriptaddress,amsmath,amssymb,nofootinbib]{revtex4-1}
\usepackage[T1]{fontenc}
\usepackage[utf8]{inputenc}
\usepackage{units}
\usepackage{amstext}
\usepackage{graphicx}
\usepackage[unicode=true,pdfusetitle,
 bookmarks=true,bookmarksnumbered=false,bookmarksopen=false,
 breaklinks=false,pdfborder={0 0 1},backref=false,colorlinks=false]
 {hyperref}
\usepackage{breakurl}

\makeatletter

\providecommand{\tabularnewline}{\\}

 \@ifundefined{textcolor}{}
 {%
   \definecolor{BLACK}{gray}{0}
   \definecolor{WHITE}{gray}{1}
   \definecolor{RED}{rgb}{1,0,0}
   \definecolor{GREEN}{rgb}{0,1,0}
   \definecolor{BLUE}{rgb}{0,0,1}
   \definecolor{CYAN}{cmyk}{1,0,0,0}
   \definecolor{MAGENTA}{cmyk}{0,1,0,0}
   \definecolor{YELLOW}{cmyk}{0,0,1,0}
 }

\usepackage{braket}

\makeatother

\begin{document}

\title{Electronic and nuclear contributions in sub-GeV dark matter scattering:
A case study with hydrogen}

\author{Jiunn-Wei Chen}

\affiliation{Department of Physics, National Taiwan University, Taipei 10617,
Taiwan}

\affiliation{Center for Theoretical Sciences and Leung Center for Cosmology and
Particle Astrophysics, National Taiwan University, Taipei 10617, Taiwan}

\affiliation{Helmholtz-Institut für Strahlen- und Kernphysik and Bethe Center
for Theoretical Physics, Universität Bonn, D-53115 Bonn, Germany}

\author{Hsin-Chang Chi}

\affiliation{Department of Physics, National Dong Hwa University, Shoufeng, Hualien
97401, Taiwan}

\author{C.-P. Liu}

\affiliation{Department of Physics, National Dong Hwa University, Shoufeng, Hualien
97401, Taiwan}

\author{Chih-Liang Wu}

\affiliation{Department of Physics, National Taiwan University, Taipei 10617,
Taiwan}

\author{Chih-Pan Wu}

\affiliation{Department of Physics, National Taiwan University, Taipei 10617,
Taiwan}

\preprint{NCTS-ECP/1501}

\begin{abstract}
Scattering of sub-GeV dark matter (DM) particles with hydrogen atoms
is studied in this paper. The interactions of DM with electrons and
nucleons are both included and formulated in a general framework based
on nonrelativistic effective field theory. On the assumption of same
dark matter coupling strengths, it is found that DM--electron interactions
dominate the inelastic atomic transitions to discrete excited states
and ionization continuum around the threshold regions, and DM--nucleon
interactions become more important with increasing energy and dominate
in elastic scattering. The conclusion should apply, qualitatively,
to practical detector species so that electronic and nuclear contributions
in DM scattering processes can be disentangled, while issues including
binding effects and recoil mechanism in many-body systems call for more 
detailed calculations.
\end{abstract}
\maketitle

\section{Introduction}

The existence of dark matter (DM) has been concretely established
based on gravitational evidences from various scales of the universe.
However, its composition and non-gravitational interactions, if any,
are still unknown. Among all DM candidates, the weakly-interacting
massive particles (WIMPs) receive most attention mainly because of
the so-called ``WIMP miracle'': with DM particle mass $m_{\chi}\sim10\,\textrm{GeV}\textrm{--}1\,\textrm{TeV}$
and self-annihilation cross section similar to the one of the weak
interaction, WIMPs can explain the relic DM abundance of the universe.
By looking for nuclear recoil triggered by WIMP scattering in detectors,
null results of direct searches have ruled out quite some WIMP parameter
space (for current status, see, e.g., \cite{Agashe:2014kda} and references
therein), and future multi-ton experiments will further improve the
limit (or make discovery).

On the other hand, the dark sector consisting of light DM (LDM) particles
with masses below 10 GeV also attracts interests recently (for overview,
see, e.g., \cite{Essig:2013lka} and references therein). First of
all, there is no reason to exclude such possibilities a priori, and
in fact, many well-motivated models predict their existence: for example,
WIMPless~\cite{Feng:2008ya,Feng:2009mn} and asymmetric DM with $m_{\chi}$
varying between MeV to GeV scales, several proposals at MeV ranges~\cite{Boehm:2003hm,Boehm:2003ha,Borodatchenkova:2005ct,Pospelov:2007mp,Fayet:2007ua,Hooper:2008im},
and keV ranges including bosonic super-WIMP~\cite{Pospelov:2008jk},
axinos~\cite{Rajagopal:1990yx,Covi:1999ty,Choi:2011yf}, and sterile
neutrinos~\cite{Dodelson:1993je,Shi:1998km,Dolgov:2000ew,Boyarsky:2008mt,Abazajian:2014gza}.
Furthermore, in order to accommodate low-energy anomalies in astrophysical
sources such as the 511 keV~\cite{Knodlseder:2005yq,Weidenspointner:2008zz}
(reviewed in \cite{Prantzos:2010wi}) and 3.5 keV~\cite{Bulbul:2014sua,Boyarsky:2014jta}
emission lines, and the GeV $\gamma$-ray excesses in the Galactic
Center~\cite{Hooper:2010mq,Hooper:2011ti}, LDM are often proposed
as possible answers.

As most of these LDM candidates can not produce observable nuclear
recoils in present-day (and near-future) detectors, direct searches
for them have to rely on electron recoils. Not like WIMP searches,
they are used to constrain DM--electron interactions~\cite{Essig:2011nj,Essig:2012yx}.
Alternative searches through indirect signals and colliders are also
reported~\cite{Essig:2013goa,Essig:2013vha,Batell:2014mga}. 

DM--electron interactions are interesting from another aspect beyond
LDM: To reconcile the tension between the light WIMP signal ($m_{\chi}\sim\textrm{10\,GeV}$)
reported by the DAMA/LIBRA experiment~\cite{Bernabei:2008yi,Bernabei:2010mq,Bernabei:2013xsa}
and stringent constraints set by other direct experiments on WIMP--nucleon
interactions, models of DM being lepto-philic but not hadro-philic
provide viable solutions (see, e.g., \cite{Bernabei:2007gr,Kopp:2009et,Foot:2014xwa}).
At even higher masses, the annihilation of WIMPs are among the favored
explanations of the positron excesses observed by PAMELA~\cite{Adriani:2008zr,Adriani:2013uda},
Fermi-LAT~\cite{FermiLAT:2011ab}, and AMS-02~\cite{Accardo:2014lma,Aguilar:2014mma}
experiments. 

If DM interacts both with electrons and nucleons, an important question
naturally arises: In a specific DM scattering process which a direct
search detector is build to look for, what are the contributions from
the electronic and nuclear degrees of freedom? Even though the current
common practice in constraining DM interactions is one-type-at-a-time,
it is necessary to keep in mind that events measured by a detector
are a sum from all possible sources. Furthermore, it is desirable
from the experimental point of view to determine which process and
kinematic region would be best to constrain a certain type of DM interactions
with electrons or nucleons. For this purpose, one has to rely on theoretical
analysis. 

In this article, we attempt to address the above questions using the simplest
atom: hydrogen -- where most calculations can be carried out analytically
-- and study its scattering with nonrelativistic LDM particles of
a MeV--GeV mass range. As the associated energy and momentum transfer
of such scattering processes overlap with typical atomic scales, one
expects that atomic physics plays an important role and issues like
binding effects and electron/nuclear recoils require detailed study.
The simplistic setup of hydrogen should therefore provide useful qualitative
understanding which applies to more intricate systems. 

The paper is organized as follows. In Section \ref{sec:formalism},
the general formalism is developed for scattering channels including
elastic, discrete excitation, and ionization. The DM--electron and
DM--nucleon interactions are formulated in a general way based on
effective field theory (EFT). In addition to commonly-discussed spin-independent
and spin-dependent contact interactions, which are the leading-order
terms in the EFT expansion, we also include the possibility of long-ranged
DM interactions (for example, through kinetic mixing of dark and normal
photons) and a few next-to-leading-order terms. In Section \ref{sec:formalism},
we present and discuss our results, and infer from these generic features
that will apply to other atoms including those practically being used
in mainstream DM detectors. Finally, a summary is given in Section \ref{sec:summary}.

\section{Formalism \label{sec:formalism}}

Direct searches of DM look for signals as results of DM scattering
off normal matter. As the nature of DM and its non-gravitational interactions
with normal matter are still unknown, instead of considering specific,
well-motivated models, we adopt a general approach based on effective
field theory (EFT). A nonrelativistic (NR) EFT that accommodates scalar,
fermionic, and vector DM particles with velocity $v_{\chi}\ll1$~\footnote{We adopt the natural units $c=1$ and $\hbar=1$.}
and their interactions with protons ($p$) and neutrons ($n$) via
intermediate scalar and vector bosons is formulated in Ref.~\cite{Fan:2010gt},
and fully worked out to next-to-next-to-leading order in Ref.~\cite{Fitzpatrick:2012ix}.
In this work, we further take electrons into account, and focus on
the kinematic region where electrons also behave like NR particles. 

At leading order (LO), the effective interaction takes the form 

\begin{align}
\mathcal{L_{\mathrm{int}}^{\mathrm{(LO)}}=} & \sum_{f=e,p,n}\left\{ c_{1}^{(f)}(\chi^{\dagger}\chi)(f^{\dagger}f)+c_{4}^{(f)}(\chi^{\dagger}\vec{S}_{\chi}\chi)\cdot(f^{\dagger}\vec{S}_{f}f)\right.\nonumber \\
 & \left.+d_{1}^{(f)}\frac{1}{q^{2}}(\chi^{\dagger}\chi)(f^{\dagger}f)+d_{4}^{(f)}\frac{1}{q^{2}}(\chi^{\dagger}\vec{S}_{\chi}\chi)\cdot(f^{\dagger}\vec{S}_{f}f)\right\} \,,\label{eq:L_LO}
\end{align}
where $\chi$ and $f$ denote the NR DM and fermion fields, respectively;
$\vec{S}_{\chi,f}$ are their spin operators (scalar DM particles
have null $\vec{S}_{\chi}$); the magnitude of the DM 3-momentum transfer
$q=\left|\vec{q}\right|$ depends on the DM energy transfer $T$ and
its scattering angle $\theta$. Note that we use the same nomenclature
as in \cite{Fitzpatrick:2012ix} for the low-energy constants (LECs)
$c_{i}^{(f)}$'s that characterize the types of the $\chi$--$f$
contact interactions. Correspondingly the LECs $d_{i}^{(f)}$'s are
used to describe potential $U(1)$-like, long-ranged $\chi$--$f$
interactions that are results of, e.g., mixing of dark and ordinary
photons via $\epsilon F_{\mu\nu}^{'}F^{\mu\nu}$ where the $F_{\mu\nu}^{^{'}}$
and $F_{\mu\nu}$ refer to the field tensors of dark and ordinary
photon, respectively, and $\epsilon$ the mixing angle. These LECs
corresponds to the ones of \cite{Fan:2010gt} by $c_{1}\rightarrow h_{1}$,
$c_{4}\rightarrow h_{2}$, $d_{1}\rightarrow l_{1}$, and $d_{4}\rightarrow l_{2}$. 

To simplify the presentation of the full scattering formula for an
unpolarized DM scattering off a hydrogen atom, we start with the simplified
case where only one of the LECs is assumed to be nonzero. (Note. This
is the conventional practice in DM searches). Following the standard
scattering theory (for more details, see, e.g., Ref.~\cite{Chen:2013iud}),
the differential cross section in the laboratory frame for DM being
scattered by the $c_{1}^{(e)}$ term alone in Eq.~(\ref{eq:L_LO})
into the final 3-momentum $\vec{k}_{2}$ with an infinitesimal phase
volume $d^{3}k_{2}$ is expressed by

\begin{align}
\left.d\sigma\right|_{c_{1}^{(e)}} & =\frac{2\pi}{v_{\chi}}\sum_{F}\overline{\sum_{I}}|\braket{F|c_{1}^{(e)}e^{i\frac{\mu}{m_{e}}\vec{q}\cdot\vec{r}}|I}|^{2}\delta(T-E_{\mathrm{CM}}-(E_{F}-E_{I}))\frac{d^{3}k_{2}}{(2\pi)^{3}}\,.
\end{align}
The reduced mass $\mu=m_{e}m_{p}/(m_{e}+m_{p})$ with $m_{e(p)}$
being the mass of electron (proton); for later use, the mass of hydrogen
is designated $M_{\mathrm{H}}=M-B$ with $M=m_{e}+m_{p}$ and $B$
the binding energy. The initial state $\ket{I}$ denotes the hydrogen
atom at the ground state, i.e., the spatial part $\ket{I}_{\mathrm{spatial}}=\ket{1s}$.
The Dirac delta function imposes the energy conservation that the
energy deposited by DM equals to the recoil energy of the atomic center
of mass, $E_{\mathrm{CM}}$, plus the internal excitation energy $E_{F}-E_{I}$
of the atom. 

Depending on the nature of the final state $\bra{F}$, the scattering
processes are classified as 
\begin{enumerate}
\item elastic scattering (el): $_{\mathrm{spatial}}\bra{F}=\bra{1s}$ ;
$E_{\mathrm{CM}}=q^{2}/(2M_{\mathrm{H}})$ and $E_{F}-E_{I}=0$,
\item discrete excitation (ex), $_{\mathrm{spatial}}\bra{F}=\bra{nlm_{l}}$
with $(n,l,m_{l})\neq(1,0,0)$; $E_{\mathrm{CM}}=q^{2}/(2M_{\mathrm{H}})$
and $E_{F}-E_{I}=E_{nl}-E_{1s}$,
\item ionization (ion): $_{\mathrm{spatial}}\bra{F}=\bra{\vec{p}_{r}}$
with $\vec{p}_{r}$ denoting the relative momentum in the CM frame;
$E_{\mathrm{CM}}=q^{2}/(2M)$ and $E_{F}-E_{I}=B+p_{r}^{2}/(2\mu)$.
\end{enumerate}
The symbol $\overline{\sum}_{I}$ means an average over the initial
magnetic (and spin, when spin operators are involved) states; $\sum_{F}$
means a sum over all the final magnetic and spin (also spin, too)
states for elastic scattering and discrete excitation, while for ionization,
the sum over magnetic states is replaced by $\int d^{3}\vec{p}_{r}/(2\pi)^{3}$. 

The analytic forms of discrete and continuum hydrogenic wave functions: 
\begin{align}
\langle100|\vec{r}\rangle= & \dfrac{1}{\sqrt{\pi}}Z^{\frac{3}{2}}e^{-Z\bar{r}}\,,\\
\langle nlm_{l}|\vec{r}\rangle= & \frac{1}{(2l+1)!}\sqrt{\frac{(n+l)!}{2n(n-l-1)!}}\left(\frac{2Z}{n}\right)^{\frac{3}{2}}e^{-\frac{Z\bar{r}}{n}}\left(\frac{2Z\bar{r}}{n}\right)^{l}{_{1}F_{1}}\left(-(n-l-1),2l+2,\frac{2Z\bar{r}}{n}\right)\nonumber \\
 & \times Y_{l}^{m_{l}*}(\theta,\phi)\,,\\
\langle\vec{p}_{r}|\vec{r}\rangle= & e^{\frac{\pi Z}{2\bar{p}_{r}}}\Gamma\left(1-\frac{iZ}{\bar{p}_{r}}\right)e^{-i\vec{p_{r}}\cdot\vec{r}}{_{1}F_{1}}\left(\frac{iZ}{\bar{p}_{r}},1,i(p_{r}r+\vec{p_{r}}\cdot\vec{r})\right)\,,
\end{align}
are given in atomic units {[}so that barred quantities $\bar{r}=rm_{e}\alpha$
and $\bar{p}_{r}=p_{r}/(m_{e}\alpha)${]}, where $\Gamma(z)$ and
$_{1}F_{1}(a,b,z)$ are the Gamma and confluent hypergeometric functions,
respectively. By the Nordsieck integration techniques~\cite{Nordsieck:1953aa,Holt:1969ar,Belkic:1981dz,Gravielle:1991mi}, we can calculate matrix elements of the transition operator 
$e^{i \vec{\kappa} \cdot \vec{r}}$, where $\vec{\kappa}$ denotes the three momentum transfer, analytically. The response function relevant for transitions to discrete states
is found to be 
\begin{align}
R^{(nl)}(\kappa) & =\sum_{m_{l}}|\braket{nlm_{l}|e^{i\vec{\kappa}\cdot\vec{r}}|1s}|^{2}\nonumber \\
 & =(2l+1)\mathcal{I}_{nl}^{2}\,,\\
\mathcal{I}_{nl}(\kappa) & =\frac{(-1)^{n-l-1}}{4n^{2}(2l+1)!}\sqrt{\frac{\pi(n+l)!}{(n-l-1)!}}\frac{\Gamma(2l+2)}{\Gamma(n+l+1)}\frac{\Gamma(2l+3)}{\Gamma(l+3/2)}(\frac{\bar{\kappa}}{4Z})^{l}\nonumber \\
 & \times\left.\left(\frac{d}{dt}\right)^{n-l-1}\left[(1-t)^{n+l+1}\left((1-t)^{2}+(\frac{\bar{\kappa}}{2Z})^{2}\right)^{-l-2}\right]\right|_{t=0}\,,
\end{align}
which is dimensionless. The response function relevant for transitions
to continuum is 
\begin{align}
R^{(ion)}(\kappa) & =\int d^{3}p_{r}|\braket{\vec{p}_{r}|e^{i\vec{\kappa}\cdot\vec{r}}|1s}|^{2}\delta(T-B-\frac{\vec{q}^{2}}{2M}-\frac{\vec{p}_{r}^{2}}{2\mu})\nonumber \\
 & =\left.\dfrac{2^{8}Z^{6}\bar{q}^{2}(3\bar{\kappa}^{2}+\bar{p}_{r}^{2}+Z^{2})\exp\left[-\dfrac{2Z}{\bar{p}_{r}}\tan^{-1}\left(\dfrac{2Z\bar{p}_{r}}{\bar{\kappa}^{2}-\bar{p}_{r}^{2}+Z^{2}}\right)\right]}{3m_{e}\alpha^{2}((\bar{\kappa}+\bar{p}_{r})^{2}+Z^{2})^{3}((\bar{\kappa}-\bar{p}_{r})^{2}+Z^{2})^{3}(1-\exp^{\frac{-2\pi Z}{\bar{p}_{r}}})}\right|_{p_{r}=\sqrt{2\mu\left(T-B-\frac{\vec{q}^{2}}{2M}\right)}}\,,
\end{align}
where the factor of $m_{e}\alpha^{2}$ in the denominator gives the
dimension as the energy conserving delta function is included in the
definition.

Using the generic response functions obtained above, the single differential
cross section with respect to energy transfer, $d\sigma/dT$, can
be compactly expressed. For elastic scattering or discrete excitation
to the final discrete level $(nl)$
\begin{align}
\left.\frac{d\sigma^{(nl)}}{dT}\right|_{c_{1}^{(e)}} & =\frac{1}{2\pi}\frac{m_{\mathrm{H}}}{v_{\chi}^{2}}\left|c_{1}^{(e)}\right|^{2}R^{(nl)}(\kappa=\frac{\mu}{m_{e}}q)\,,\label{eq:dS/dT_nl}\\
 & \mathrm{with}\quad q^{2}=2M_{\mathrm{H}}(T-(E_{nl}-E_{1s}))\,.\label{eq:q^2_discrete}
\end{align}
The $1/v_{\chi}^{2}$ factor in Eq.~(\ref{eq:dS/dT_nl}) comes from
two sources with each one contributing $1/v_{\chi}$: (i) division
by flux in $d\sigma/dT$ and (ii) the integration of DM scattering
angle $\cos\theta$ with respect to the energy conserving delta function.
Note that it does not lead to physical sigularity when taking an extremely
NR limit $v_{\chi}\rightarrow0$, since the DM flux and kinetic energy
both approach zero. The magnitude of $q$ is determined by energy conservation, or equivalently,
the scattering angle $\cos\theta$ is fixed once the energy transfer
$T$ for such 2-to-2-body scattering is known. For ionization processes: 

\begin{align}
\left.\frac{d\sigma^{(ion)}}{dT}\right|_{c_{1}^{(e)}} & =\frac{1}{2\pi}\frac{m_{\chi}}{v_{\chi}}k_{2}\int d\cos\theta\,\left|c_{1}^{(e)}\right|^{2}R^{(ion)}(\kappa=\frac{\mu}{m_{e}}q),\label{eq:dS/dT_pr}\\
 & \mathrm{with}\quad\min\left\{ 1,\max\left[-1,\dfrac{k_{1}^{2}+k_{2}^{2}-2M(T-B)}{2k_{1}k_{2}}\right]\right\} \leq\cos\theta\leq1\,,\label{eq:q^2_ion}
\end{align}
where $k_{1}=m_{\chi}v_{\chi}$ and $k_{2}=(m_{\chi}^{2}v_{\chi}^{2}-2m_{\chi}T)^{1/2}$
are the magnitudes of the initial and final momentum, respectively.
Because the final atomic state has two bodies to share the transferred
energy and momentum, the DM scattering angle $\cos\theta$ now can
span a finite range for a given energy transfer $T$. 

Next we consider the $d_{1}^{(e)}$ term. Because its Lagrangian differs
from $c_{1}^{(e)}$ term only by a kinematic factor $1/q^{2}$ (which
only cause a rescaling of transition matrix elements), it can easily
be calculated by replacing $c_{1}^{(e)}$ to $d_{1}^{(e)}/q^{2}$
in both Eqs.~(\ref{eq:dS/dT_nl}) and (\ref{eq:dS/dT_pr}). If both
terms exist, then one has to take their coherent interference into
account, so that the coupling in front of the response functions becomes
$c_{1}^{(e)}+d_{1}^{(e)}/q^{2}$.

Unlike the $c_{1}^{(e)}$ and $d_{1}^{(e)}$ terms, which are independent
of the spins of the DM and the scattered particles, the $c_{4}^{(e)}$
and $d_{4}^{(e)}$ terms give rise to what typically called spin-dependent
interactions. Their matrix elements for unpolarized scattering involve
additionally initial spin average and final spin sum. For a spinor
with spin quantum number $s$ and $m_{s}$, it yields 
\begin{align}
\sum_{m_{s}^{'}} & \overline{\sum_{m_{s}}}\braket{s,m_{s}^{'}|S_{_{a}}|s,m_{s}}\braket{s,m_{s}^{'}|S_{_{b}}|s,m_{s}}^{*}=\frac{1}{3}s(s+1)\delta_{ab}\,.\label{eq:spin_mat}
\end{align}
With the DM spin $s_{\chi}$ and the electron spin $s_{e}=1/2$, the
spin averages and sums applied to both the DM and electron parts yield
a product: $s_{\chi}(s_{\chi}+1)/4$. Other than this factor, the
rest of spatial matrix elements are completely the same as in the
$c_{1}^{(e)}$ and $d_{1}^{(e)}$ case. As a result, the corresponding
scattering formula can be obtained by changing $\left|c_{1}^{(e)}+d_{1}^{(e)}/q^{2}\right|^{2}$
to $\frac{1}{4}s_{\chi}(s_{\chi}+1)\left|c_{4}^{(e)}+d_{4}^{(e)}/q^{2}\right|^{2}$
in both Eqs.~(\ref{eq:dS/dT_nl}) and (\ref{eq:dS/dT_pr}). 

It is worthwhile to point out here that there is no interference between
the spin-independent interactions with $c_{1}$, $d_{1}$ and the
spin-dependent one with $c_{4}$, $d_{4}$, in unpolarized scattering,
since the trace of a spin matrix is zero.

Now we consider the cases when DM scatters off the proton instead
of the electron. Besides the trivial change of LECs, the most important
difference is due to the fact that the proton is much closer to the
atomic center of mass than the electron. After factoring out the center-of-mass
motion, the resulting atomic transition operators in its intrinsic
frame are 
\begin{align}
\rho^{(e)}(\vec{q}) & =e^{i\frac{\mu}{m_{e}}\vec{q}\cdot\vec{r}}\,,\\
\rho^{(p)}(\vec{q}) & =e^{-i\frac{\mu}{m_{p}}\vec{q}\cdot\vec{r}}\,,
\end{align}
for the electron and the proton, respectively. This leads to the following
change of the corresponding response functions: 
\begin{align}
R_{e}^{(nl,ion)} & =R^{(nl,ion)}(\kappa=\frac{\mu}{m_{e}}q)\,,\\
R_{p}^{(nl,ion)} & =R^{(nl,ion)}(\kappa=\frac{\mu}{m_{p}}q)\,,
\end{align}
and similarly in the differential cross section formulae, Eqs.~(\ref{eq:dS/dT_nl})
and (\ref{eq:dS/dT_pr}).

Finally we can summarize the above derivation and obtain the differential
cross section formulae for DM scattering off the hydrogen atom at
LO. For transitions to discrete states: 

\begin{align}
\left.\frac{d\sigma^{(nl)}}{dT}\right|_{\mathrm{LO}} & =\frac{1}{2\pi}\frac{m_{\mathrm{H}}}{v_{\chi}^{2}}\left\{ \sum_{f=e,p}\left(\left|c_{1}^{(f)}+\frac{d_{1}^{(f)}}{q^{2}}\right|^{2}+\frac{1}{4}s_{\chi}(s_{\chi}+1)\left|c_{4}^{(f)}+\frac{d_{4}^{(f)}}{q^{2}}\right|^{2}\right)R_{f}^{(nl)}\right.\nonumber \\
 & \left.\left.+2\left(c_{1}^{(e)}+d_{1}^{(e)}/q^{2}\right)\left(c_{1}^{(p)}+d_{1}^{(p)}/q^{2}\right)^{*}R_{ep}^{(nl)}\right\} \right|_{q^{2}=2M_{\mathrm{H}}(T-(E_{nl}-E_{1s}))}\,,\label{eq:dS/dT_nl_LO}
\end{align}
and for ionizations: 

\begin{align}
\left.\frac{d\sigma^{(ion)}}{dT}\right|_{\mathrm{LO}} & =\frac{1}{2\pi}\frac{m_{\chi}}{v_{\chi}}k_{2}\int d\cos\theta\left\{ \sum_{f=e,p}\left(\left|c_{1}^{(f)}+\frac{d_{1}^{(f)}}{q^{2}}\right|^{2}+\frac{1}{4}s_{\chi}(s_{\chi}+1)\left|c_{4}^{(f)}+\frac{d_{4}^{(f)}}{q^{2}}\right|^{2}\right)R_{f}^{(ion)}\right.\nonumber \\
 & \left.\left.+2\left(c_{1}^{(e)}+d_{1}^{(e)}/q^{2}\right)\left(c_{1}^{(p)}+d_{1}^{(p)}/q^{2}\right)^{*}R_{ep}^{(ion)}\right\} \right|_{p_{r}^{2}=2\mu\left(T-B-q^{2}/(2M)\right)}\,.\label{eq:dS/dT_pr_LO}
\end{align}
In these formulae, there are two new response functions $R_{ep}^{(nl,ion)}$
defined, which describe the nontrivial interference between the spin-independent
$\chi$--$e$ and $\chi$--$p$ amplitude; they are

\begin{equation}
R_{ep}^{(nl)}=\sqrt{R_{e}^{(nl)}R_{p}^{(nl)}}\,,
\end{equation}
and

\begin{equation}
R_{ep}^{(ion)}=\int d^{3}p_{r}\braket{\vec{p}_{r}|e^{i\frac{\mu}{m_{e}}\vec{q}\cdot\vec{r}}|1s}\braket{\vec{p}_{r}|e^{-i\frac{\mu}{m_{p}}\vec{q}\cdot\vec{r}}|1s}^{*}\delta(T-B-\frac{\vec{q}^{2}}{2M}-\frac{\vec{p}_{r}^{2}}{2\mu})\,.
\end{equation}
Note that there is no interference between the $c_{4}^{(e)}$ and
$c_{4}^{(p)}$ terms, nor between the $d_{4}^{(e)}$ and $d_{4}^{(p)}$
terms, as they involve different spin operators with each of them
traceless.

Even though we mainly concentrate on the LO interaction with DM in
this article, we shall also consider a few terms at next-to-leading
order (NLO): 

\begin{align}
\mathcal{L_{\mathrm{int}}^{\mathrm{(NLO)}}=} & \sum_{f=e,p,n}\left\{ c_{10}^{(f)}(\chi^{\dagger}\chi)(f^{\dagger}i\vec{\sigma}_{f}\cdot\vec{q}f)+c_{11}^{(f)}(\chi^{\dagger}i\vec{\sigma}_{\chi}\cdot\vec{q}\chi)(f^{\dagger}f)\right.\nonumber \\
 & \left.+d_{10}^{(f)}\frac{1}{q^{2}}(\chi^{\dagger}\chi)(f^{\dagger}i\vec{\sigma}_{f}\cdot\vec{q}f)+d_{11}^{(f)}\frac{1}{q^{2}}(\chi^{\dagger}i\vec{\sigma}_{\chi}\cdot\vec{q}\chi)(f^{\dagger}f)\right\} +\cdots
\end{align}
They translate into the ones of \cite{Fan:2010gt} by $c_{11}\rightarrow h_{1}^{'}$,
$c_{10}\rightarrow h_{2}^{'}$, $d_{11}\rightarrow l_{1}^{'}$, and
$d_{10}\rightarrow l_{2}^{'}$. Because the spin operators that come
with the $c_{11}^{(e,p)}$ and $c_{10}^{(e,p)}$ terms are mutually
independent (after spin average and sum) with each other except for
the interference between $c_{11}^{(e)}$ and $c_{11}^{(p)}$, and
also with all LO terms, including them into Eqs.~(\ref{eq:dS/dT_nl_LO})
and (\ref{eq:dS/dT_pr_LO}) is straightforward by 

\begin{align}
 & \left(\left|c_{1}^{(f)}+\frac{d_{1}^{(f)}}{q^{2}}\right|^{2}+\frac{1}{4}s_{\chi}(s_{\chi}+1)\left|c_{4}^{(f)}+\frac{d_{4}^{(f)}}{q^{2}}\right|^{2}\right)\nonumber \\
 & \rightarrow\left(\left|c_{1}^{(f)}+\frac{d_{1}^{(f)}}{q^{2}}\right|^{2}+\frac{1}{4}s_{\chi}(s_{\chi}+1)\left|c_{4}^{(f)}+\frac{d_{4}^{(f)}}{q^{2}}\right|^{2}+\frac{1}{3}s_{\chi}(s_{\chi}+1)q^{2}\left|c_{11}^{(f)}+\frac{d_{11}^{(f)}}{q^{2}}\right|^{2}+\frac{1}{4}q^{2}\left|c_{10}^{(f)}+\frac{d_{10}^{(f)}}{q^{2}}\right|^{2}\right)\nonumber \\
 & \left(c_{1}^{(e)}+d_{1}^{(e)}/q^{2}\right)\left(c_{1}^{(p)}+d_{1}^{(p)}/q^{2}\right)^{*}\nonumber \\
 & \rightarrow\left(c_{1}^{(e)}+d_{1}^{(e)}/q^{2}\right)\left(c_{1}^{(p)}+d_{1}^{(p)}/q^{2}\right)^{*}+\frac{1}{3}s_{\chi}(s_{\chi}+1)q^{2}\left(c_{11}^{(e)}+d_{11}^{(e)}/q^{2}\right)\left(c_{11}^{(p)}+d_{11}^{(p)}/q^{2}\right)^{*}\,,
\end{align}
where similar argument is applied to the $d_{11}^{(e,p)}$ and $d_{10}^{(e,p)}$
terms.

\section{Results And Discussions \label{sec:results}}

In this section, we give numerical results for two different DM masses:
(i) $m_{\chi}=1\,\mathrm{GeV}$ and (ii) $m_{\chi}=50\,\mathrm{MeV}$,
and with a nonrelativistic velocity $v_{\chi}=10^{-3}$. Our main
purpose is to illustrate and discuss how electron and proton respectively
contribute to the scattering processes for specified EFT interaction
terms and reaction channels. For clarity in presentation, we ignore
all interference terms and assume the DM interaction strengths with
electron and proton are the same when making comparisons between the
electronic and nuclear contributions. It should be borne in mind that
the total cross section is a sum of all contributions from electron
and proton with interference terms included.

\subsection{LO Interactions of $c_{1}$, $d_{1}$, $c_{4}$, and $d_{4}$}

\begin{figure}[h]
\begin{tabular}{cc}
\includegraphics[scale=0.4]{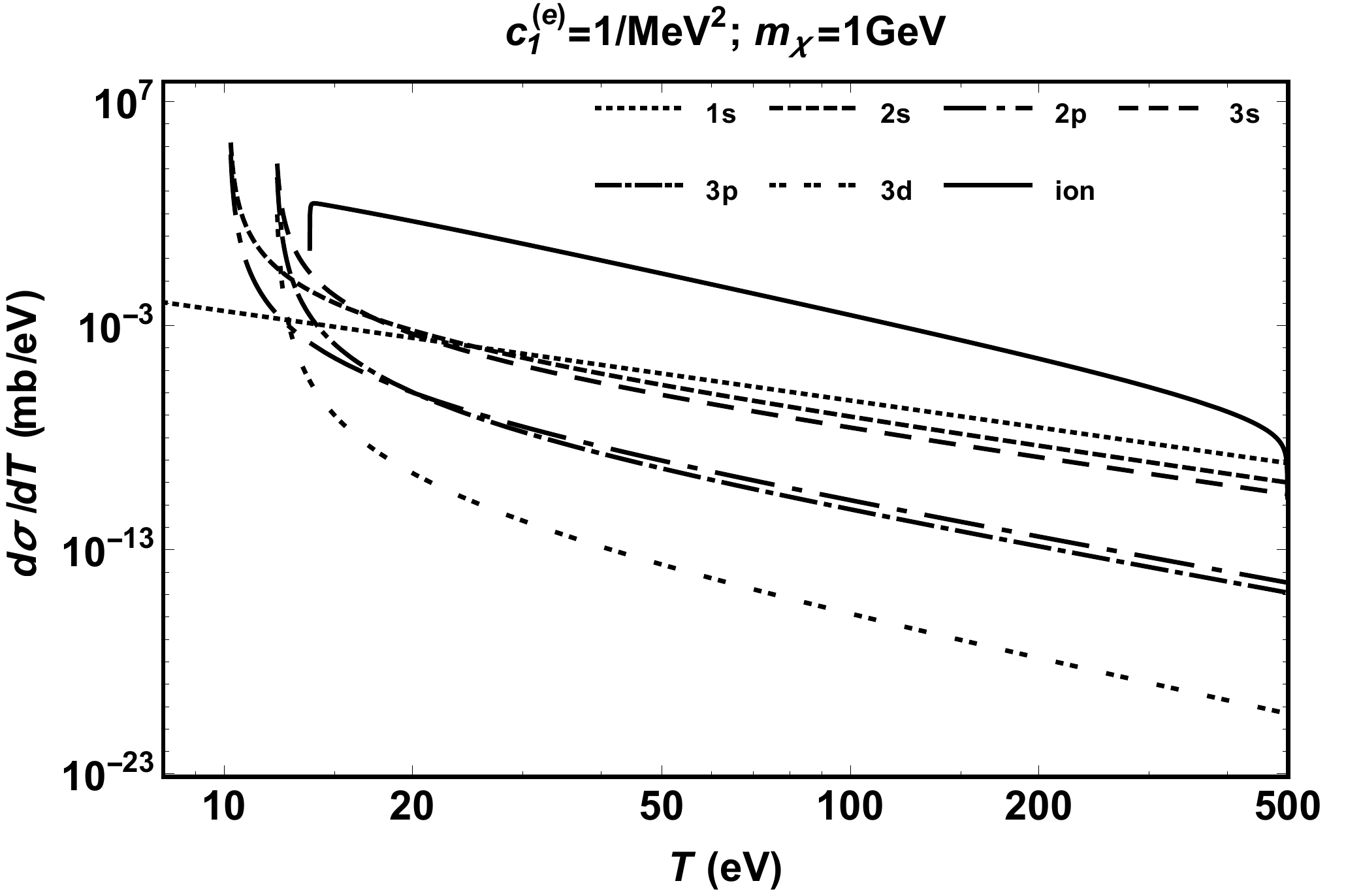} & \includegraphics[scale=0.4]{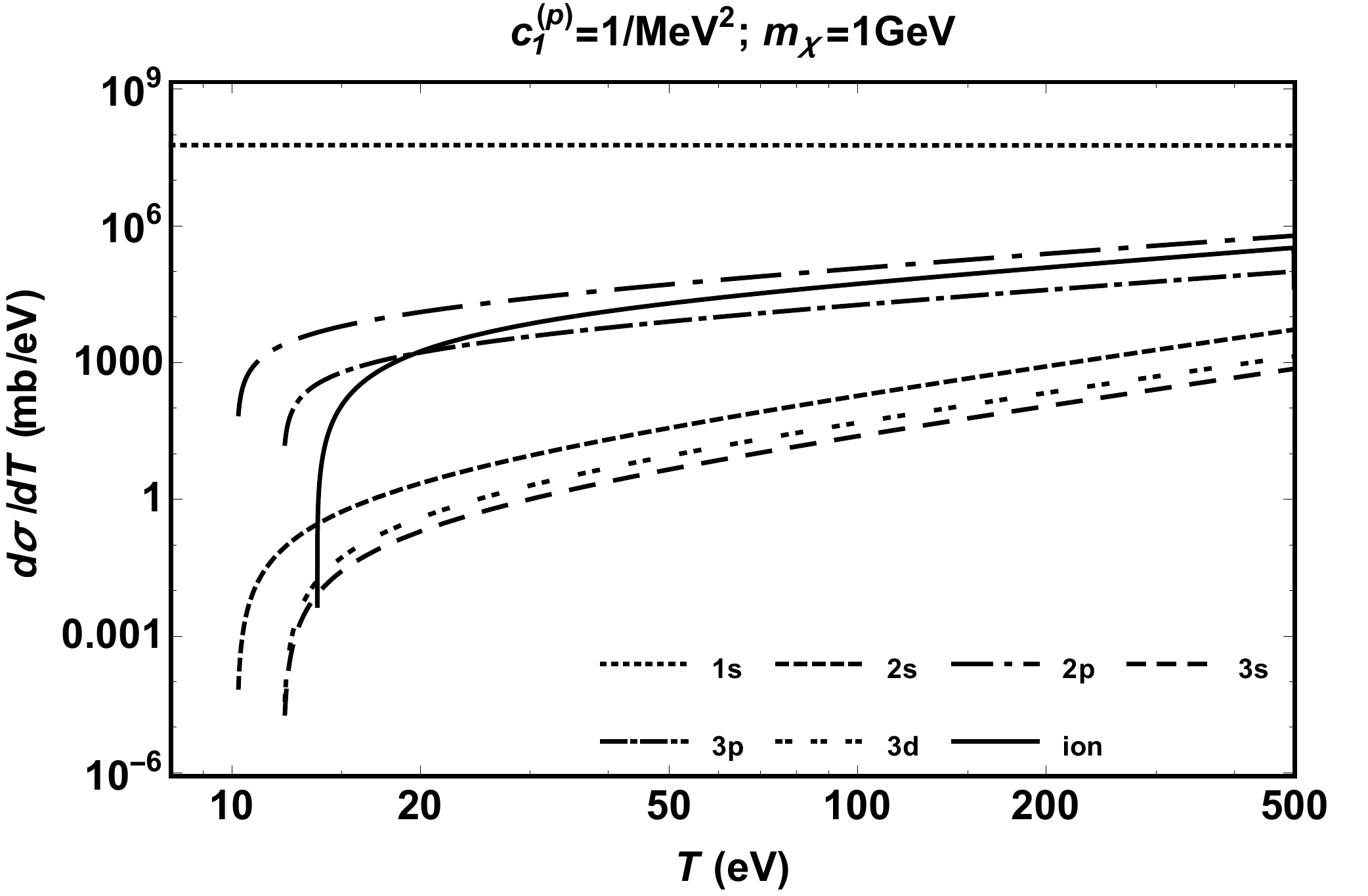}\tabularnewline
\includegraphics[scale=0.4]{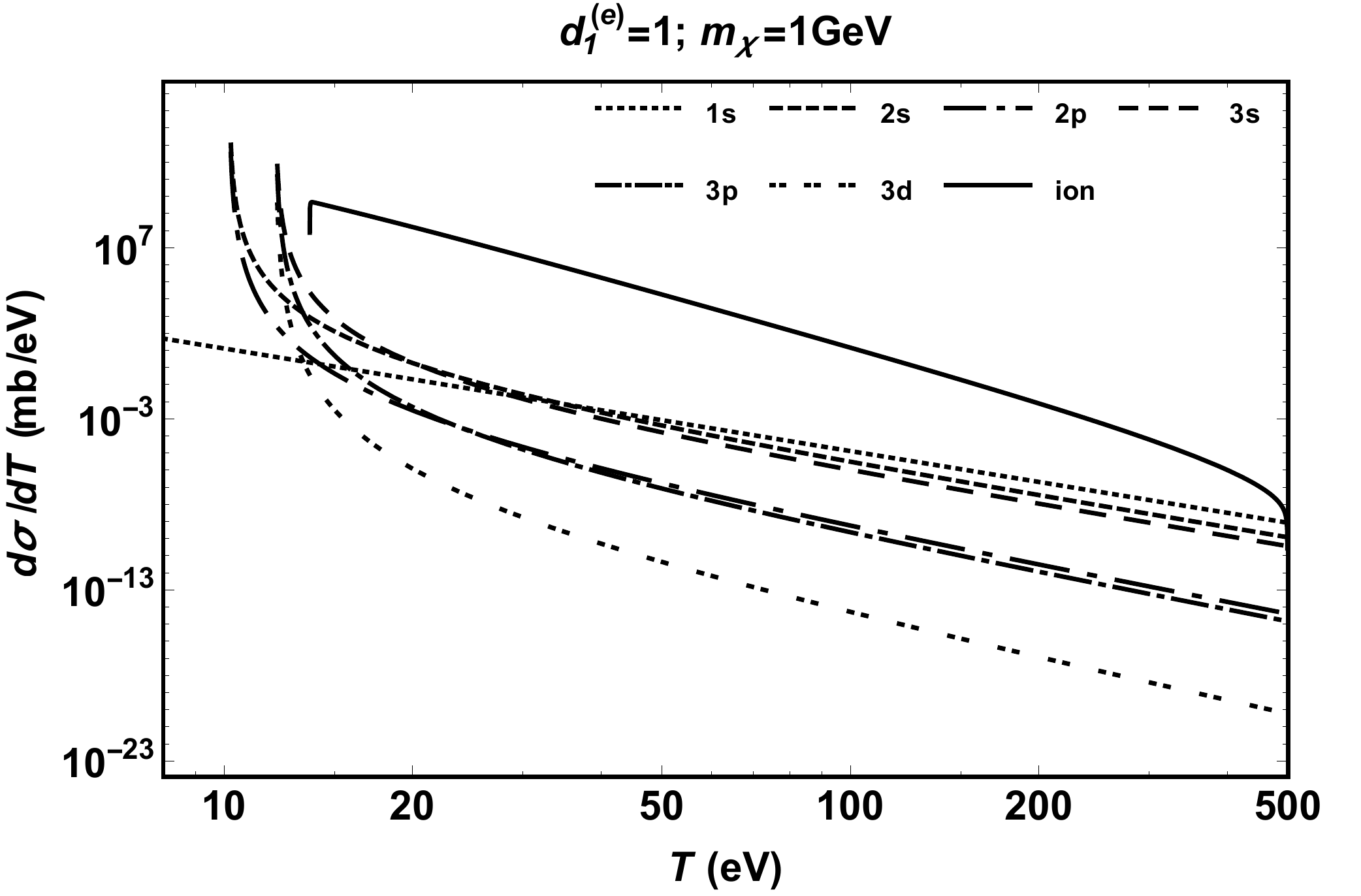} & \includegraphics[scale=0.4]{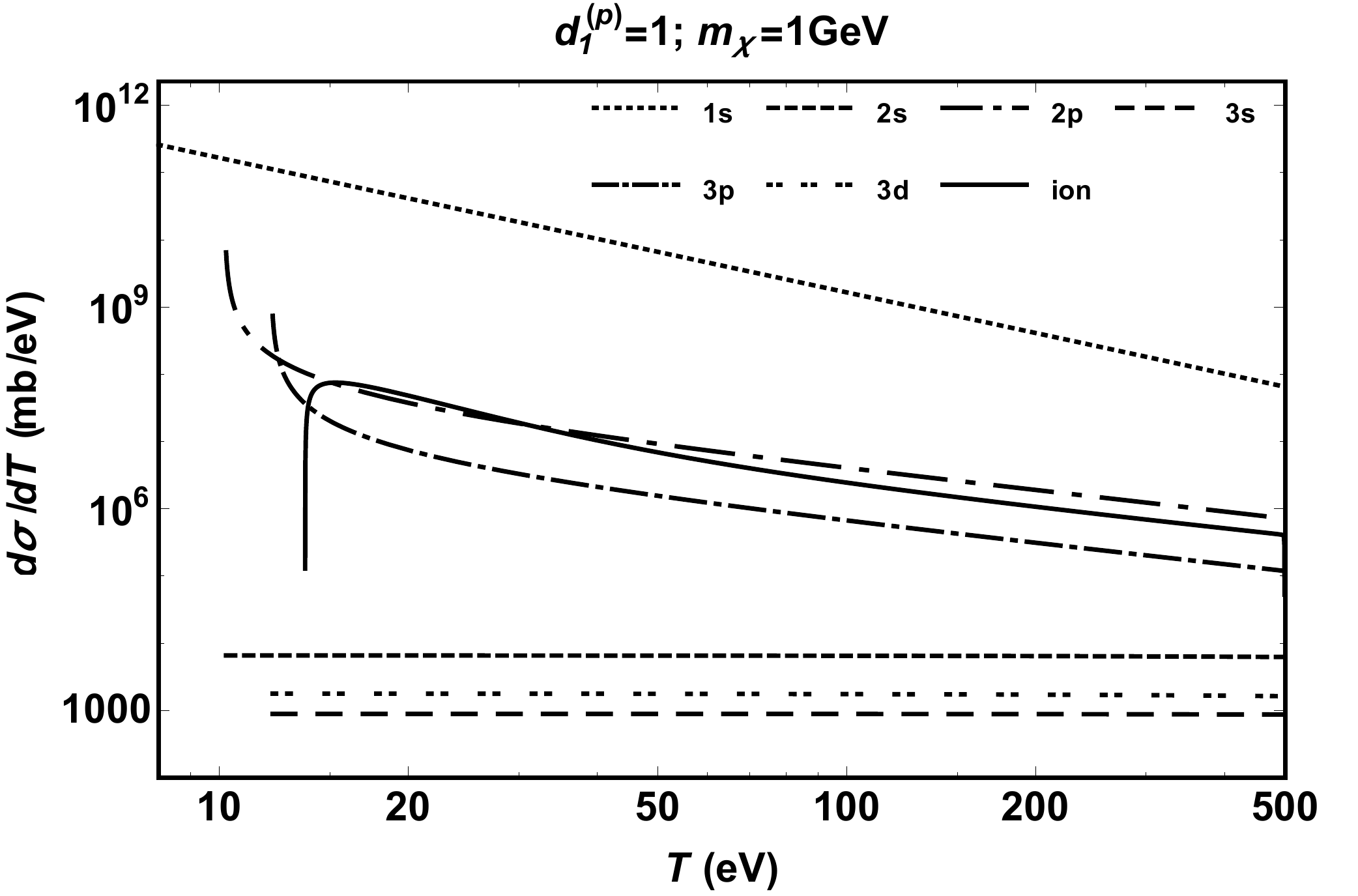}\tabularnewline
\end{tabular} 

\protect\caption{Different channels of DM scattering with $m_{\chi}=1\,\mathrm{GeV}$
and interaction of $c_{1}^{(e)}=1/\textrm{MeV}^{2}$ (upper-left),
$c_{1}^{(p)}=1/\textrm{MeV}^{2}$ (upper-right), $d_{1}^{(e)}=1$
(lower-left), $d_{1}^{(p)}=1$ (lower-right). \label{fig:c1_1GeV} }
\end{figure}

The upper two panels of Fig.~\ref{fig:c1_1GeV} show the differential
cross sections $d\sigma/dT$ for DM scattering with $m_{\chi}=1\,\mathrm{GeV}$
and the $c_{1}$-type interactions. For the nuclear part (right panel),
the elastic scattering dominates all other channels by orders of magnitude.
The reason is obvious: Since the momentum scale that determines the
nuclear response $\kappa_{p}=\nicefrac{\mu}{m_{p}}q\sim\nicefrac{\mu}{m_{p}}m_{\chi}v_{\chi}\sim0.5\,\mathrm{keV}$
is smaller than the inverse of atomic size, $m_{e}\alpha\sim3\,\mathrm{keV}$,
it is a good approximation that the nuclear charge operator can be
expanded as $\rho(\vec{\kappa}_{p})\sim1-i\vec{\kappa}_{p}\cdot\vec{r}+\ldots$
Unlike the elastic scattering, all inelastic channels have no leading-order
contributions because of wave function orthogonality. Also because
the next-leading-order operator is a dipole operator, the excitations
to final $p$-orbitals or continuum are more probable than other discrete
states. For the electronic part (left panel), the results change dramatically.
First, as the momentum scale that determines the electronic response
$\kappa_{e}=\nicefrac{\mu}{m_{e}}q\sim\nicefrac{\mu}{m_{e}}m_{\chi}v_{\chi}\sim1\,\mathrm{MeV}$
is much bigger than the inverse of atomic size, the electric charge
operator $\rho(\vec{\kappa}_{e})=e^{i\vec{\kappa}_{e}\cdot\vec{r}}$
becomes highly oscillating. As a result, the elastic differential
cross section shows the familiar form factor suppression. On the other
hand, in discrete excitations, one does observe much larger cross
sections in near-threshold regions. The reason is most of the energy
transferred $T$ by DM is given to internal excitation; this leaves
the 3-momentum transfer $q=\sqrt{2M_{\mathrm{H}}(T-(E_{nl}-E_{1s}))}$
becoming quite small so that the form factor suppression is less severe.
Among all reaction channels arising from the $c_{1}^{(e)}$-type interaction,
the ionization channel is the dominant one in most of the kinematic
region, for it can access more of the kinematic phase space with small
$q$. In addition, the peaks near discrete excitation thresholds also
provide good observation windows for not only the large cross section
but also the clean signal of deexcitation photons of definite energies. 

In the lower two panels of Fig.~\ref{fig:c1_1GeV}, the results for
the $d_{1}$-type interactions are shown. The numerical changes from
the previous $c_{1}$-type results are mainly due to adding $1/q^{2}$
factor in the scattering amplitude (or $1/q^{4}$ in the double differential
cross section). Because $q^{2}=2m_{\mathrm{H}}T$ in elastic scattering,
this leads to an extra $1/T^{2}$ dependence in $d\sigma/dT$ and
can be best illustrated by observing the difference of the (almost)
flat line and the power-law-decreasing line in the $c_{1}^{(p)}$
and $d_{1}^{(p)}$ plots respectively. For discrete excitation channels,
except for near-threshold region, one expects similar $1/T^{2}$ dependence
when $T$ gets bigger than excitation energies. The case of ionization
channel is more intricate, as $q^{2}$ is to be integrated over a
range allowed by kinematics; there is no simple scaling from the $c_{1}$-
to $d_{1}$-type results. Overall, the long-range interaction yields
sharper energy dependence of $d\sigma/dT$ than the contact one for
all scattering channels considered. The elastic scattering is still
the best channel to constrain $d_{1}^{(p)}$, and discrete excitations
at thresholds and ionization the best for $d_{1}^{(e)}$.

\begin{figure}[h]
\begin{tabular}{cc}
\includegraphics[scale=0.4]{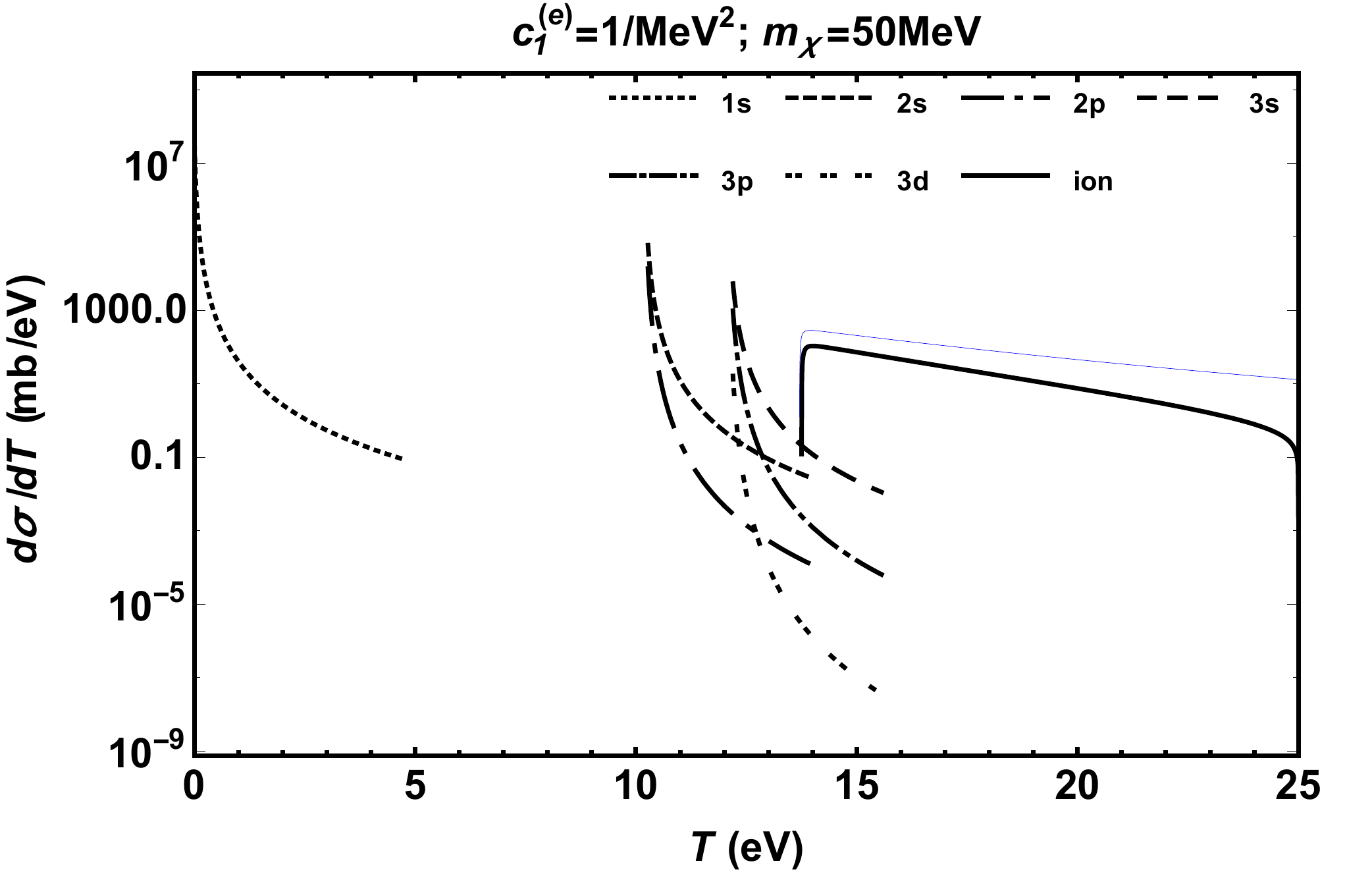} & \includegraphics[scale=0.4]{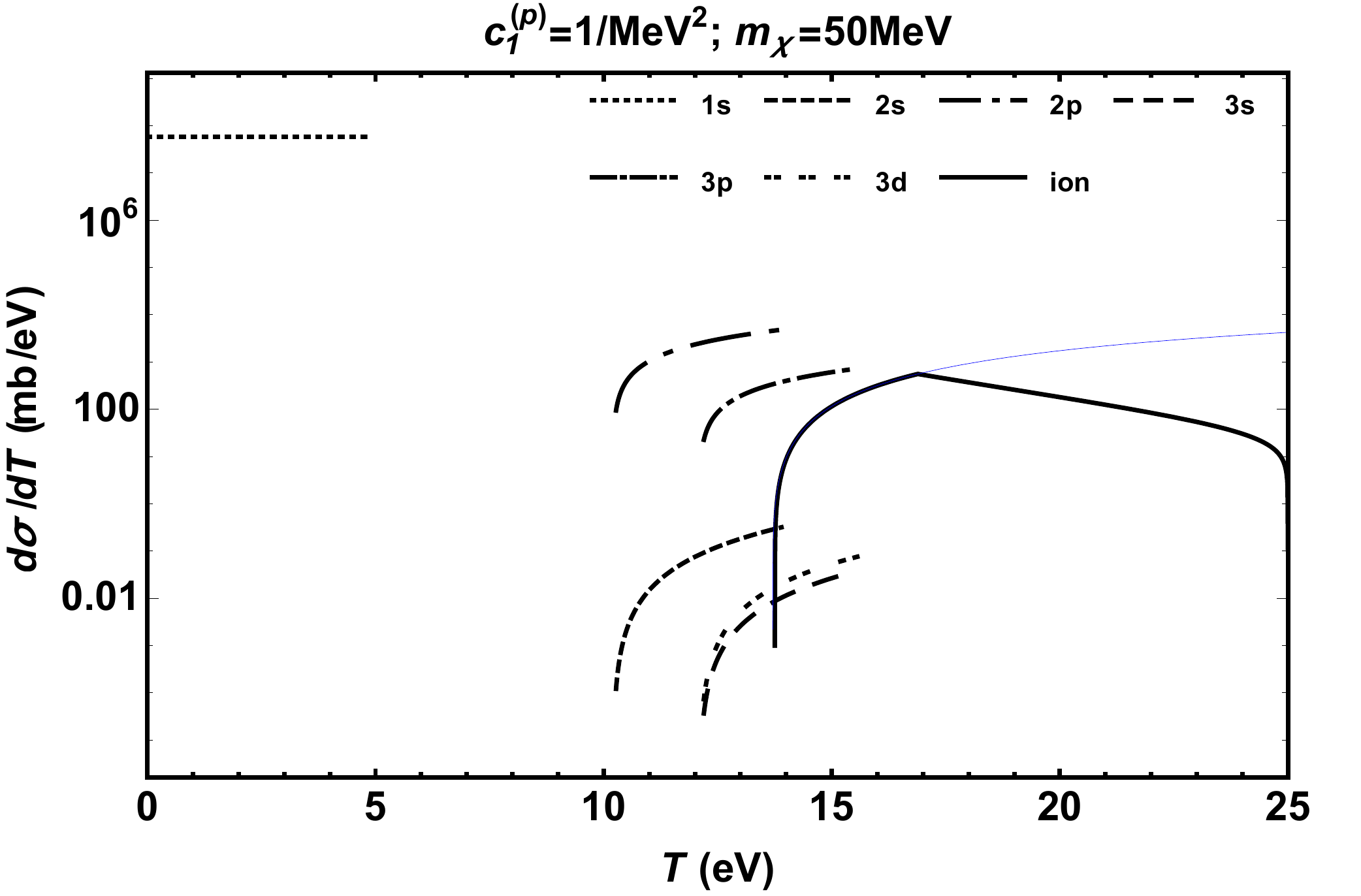}\tabularnewline
\includegraphics[scale=0.4]{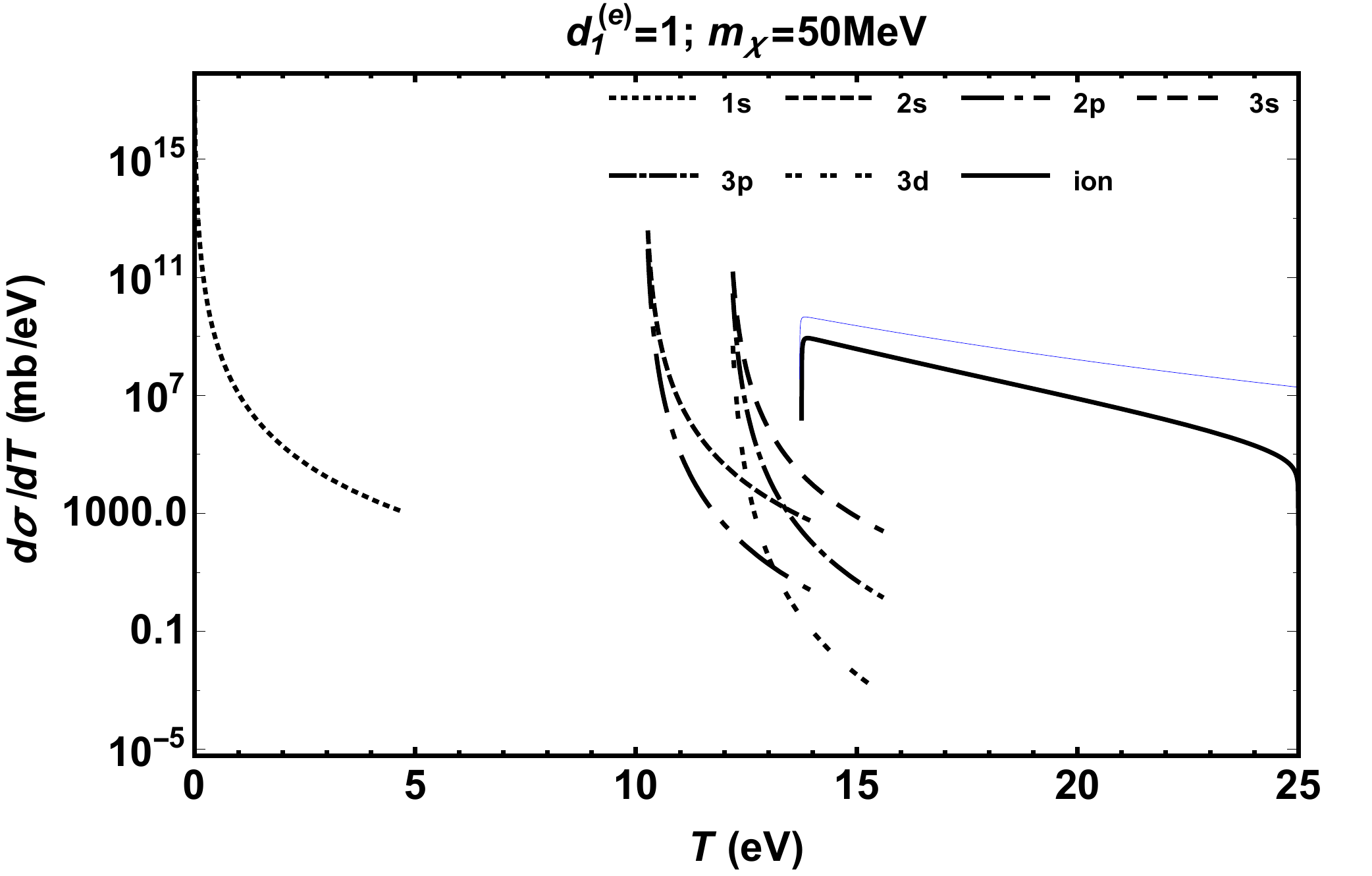} & \includegraphics[scale=0.4]{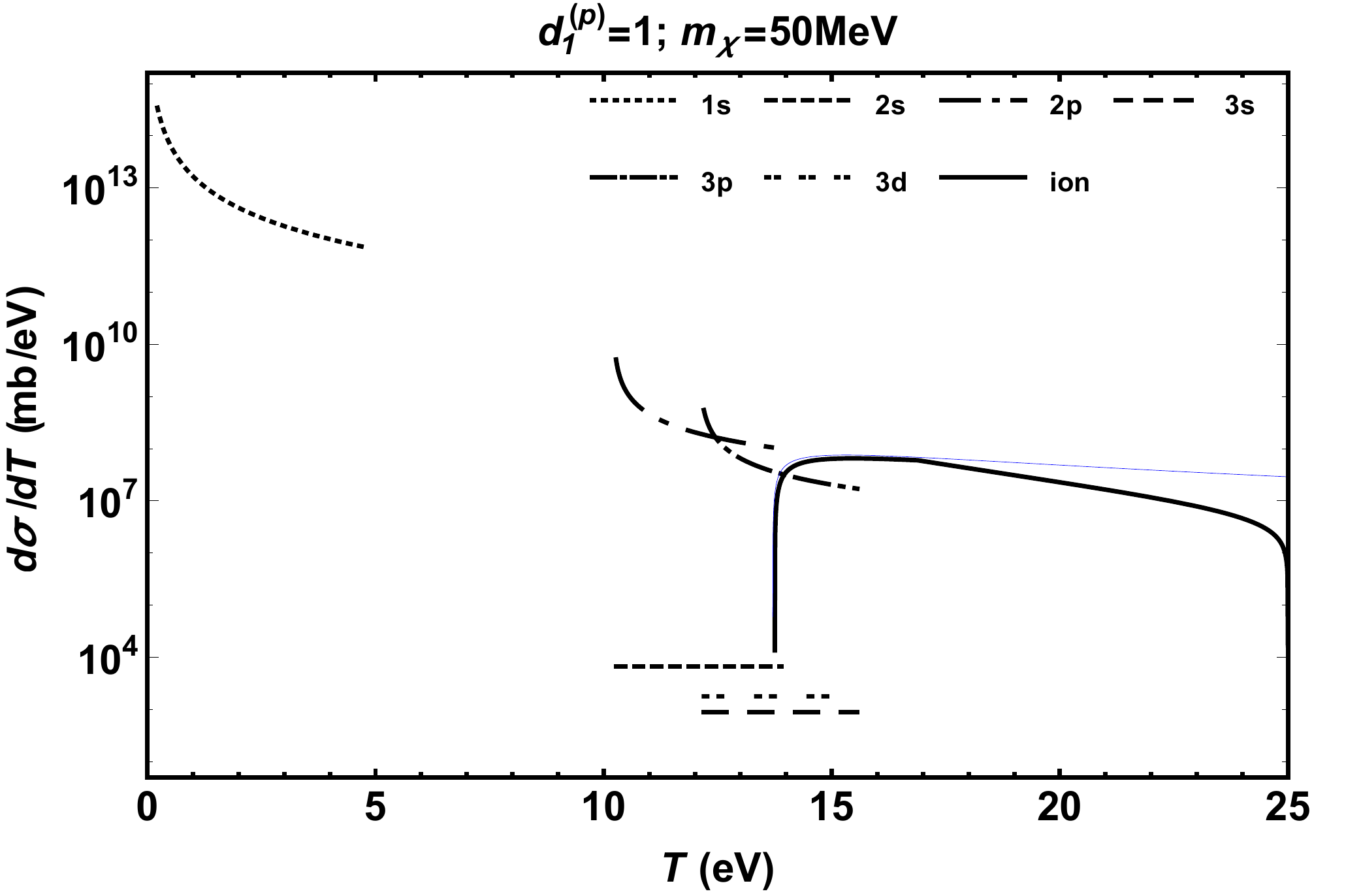}\tabularnewline
\end{tabular} 

\protect\caption{Different channels of DM scattering with $m_{\chi}=50\,\mathrm{MeV}$
and interaction of $c_{1}^{(e)}=1/\textrm{MeV}^{2}$ (upper-left),
$c_{1}^{(p)}=1/\textrm{MeV}^{2}$ (upper-right), $d_{1}^{(e)}=1$
(lower-left), $d_{1}^{(p)}=1$ (lower-right). The thin blue curves
are the ionization results with $m_{\chi}=1\,\textrm{GeV}$, shown
for comparison. (Note. Unlike Fig.~\ref{fig:c1_1GeV}, these are
linear-log plots.)\label{fig:c1_50MeV} }
\end{figure}

In Fig.~\ref{fig:c1_50MeV}, similar plots but with $m_{\chi}=50\,\mathrm{MeV}$
are shown. The most noticeable differences from Fig.~\ref{fig:c1_50MeV}
are (i) the NR DM kinetic energy is smaller so $T_{\max}=\nicefrac{1}{2}m_{\chi}v_{\chi}^{2}=25\,\textrm{eV}$
and (ii) in elastic scattering and discrete excitations, the maximum
allowed energy transfers are cut off at smaller values: 4.8, 14.0,
and 15.7 eV for final $n=1$, 2, 3, respectively. The latter is due
to maintaining energy and momentum conservation in the final two-body
system (the DM particle and the atom) with $m_{\chi}\leq M_{H}$.
Using Eq.~(\ref{eq:q^2_discrete}) and setting the maximum DM scattering
angle $\cos\theta=-1$, one can get an approximate formula 
\begin{equation}
T_{\textrm{cut}}^{(nl)}=\frac{4m_{\chi}M_{H}}{(m_{\chi}+M_{H})^{2}}T_{\max}+\frac{M_{H}-m_{\chi}}{M_{H}+m_{\chi}}(E_{nl}-E_{1s})\,,\label{eq:T_cut}
\end{equation}
which yields the correct cut off energies just pointed out. Except
the cutoffs in energy transfer, $d\sigma/dT$'s of elastic scattering
and discrete excitations are the same for both $m_{\chi}=50\,\textrm{MeV}$
and $m_{\chi}=1\,\textrm{GeV}$ in the allowed range of $T$. Because
the associated response functions, which depend on the 3-momentum
transfer $q$, are fixed by $T$ and excitation energy, the independence
of $m_{\chi}$ is thus understood. Note that these cutoff energies
limit the ability of direct LDM searches because the recoil energies
are too small to be detected.

On the other hand, $d\sigma/dT$ of the ionization channel shows different
features. First, as there are three bodies in the final states (the
DM particle, the ionized atom, and the ejected electron), energy and
momentum conservation does not introduce a kinematic cutoff so $T$
can extend to the end point energy $T_{\max}$. For this reason that
the ionization channel should be considered as the golden mode to
LDM direct searches. Second, the value of $q$ does depend on $m_{\chi}$,
via Eq.~(\ref{eq:q^2_ion}); as a result, $d\sigma/dT$ is not $m_{\chi}$
independent. To make the comparison clear, the results of $m_{\chi}=1\,\textrm{GeV}$
are plotted with thin solid curves in the same figures. 

At $T\approx17\,\mathrm{eV}$, one observes discontinuities in $d\sigma/dT$
on the right panels of Fig.~\ref{fig:c1_50MeV}. This is a combined
result of two ingredients: (i) The scattering angle is bounded by
Eq.~(\ref{eq:q^2_ion}). At $T\approx17\,\textrm{eV}$, the maximum
scattering angle $180^{\circ}$ is reached (this energy can also be
anticipated from Eq.~(\ref{eq:T_cut}) with the excitation energy
$E_{nl}-E_{1s}$ being replaced by the binding energy $-E_{1s}$ ),
so the integration range in Eq.~(\ref{eq:dS/dT_pr}) ceases to increase
for $T>17\,\textrm{eV}$. (ii) The nuclear response function $R_{p}^{(ion)}$
is bigger at backward angels than forward angels, so the integral
Eq.~(\ref{eq:dS/dT_pr}) sensitively depends on the integration range
and its discontinuity. On the contrary, the electronic response function
$R_{e}^{(ion)}$ is only significant at small angels, therefore the
discontinuity in the integration range of Eq.~(\ref{eq:dS/dT_pr})
does not yield observable results on the left panels of Fig.~\ref{fig:c1_50MeV}.

The hydrogen atom only has one electron and one nucleon, so the contributions
from the $c_{4}$- and $d_{4}$-type interactions are related to the
ones of the $c_{1}$- and $d_{1}$-type interactions simply by a rescaling
of one-body spin matrix element as discussed in the last section.
We shall not repeat these plots, but just note that for other atoms
with more electrons and nucleons, the spin-dependent cross sections
from interaction terms like $c_{4}$ and $d_{4}$ do not receive many-body
enhancement compared with the spin-independent interactions terms
like $c_{1}$ and $d_{1}$. 

\begin{figure}[h]
\begin{tabular}{cc}
\includegraphics[scale=0.4]{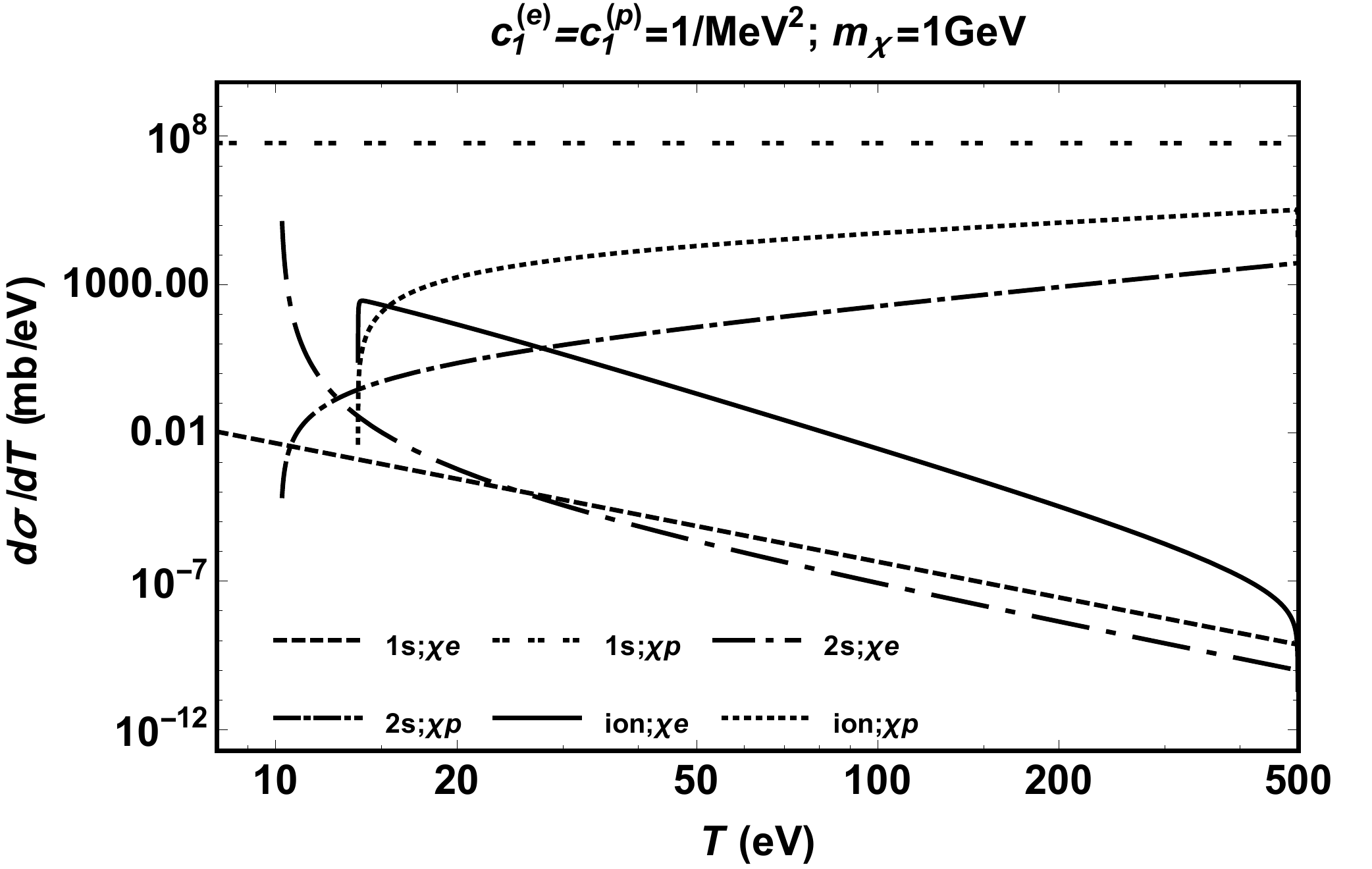} & \includegraphics[scale=0.4]{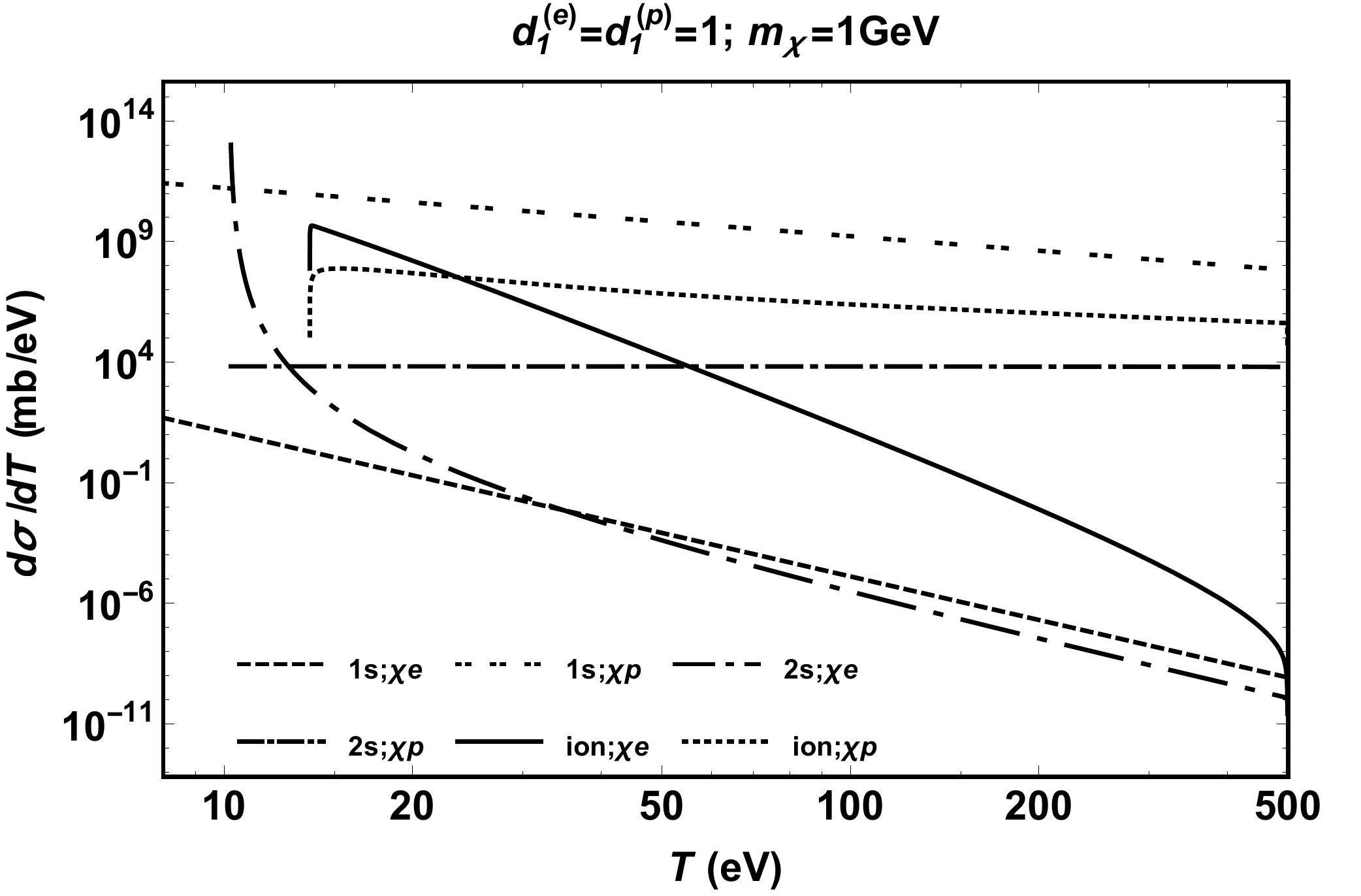}\tabularnewline
\end{tabular} 

\protect\caption{Comparison of DM cross sections with the electron ($\chi e$) and
proton ($\chi p$) in a hydrogen atom for selected channels including
(i) elastic (1s), (ii) discrete excitation to 2s, and (iii) ionization
(ion). The interactions are taken to be (Left) $c_{1}^{(e)}=c_{1}^{(p)}=1/\textrm{MeV}^{2}$
and (Right) $d_{1}^{(e)}=d_{1}^{(p)}=1$. Interference terms due to
$\chi e$ and $\chi p$ amplitudes are ignored. \label{fig:c1_comp} }
\end{figure}

\begin{figure}[h]
\begin{tabular}{cc}
\includegraphics[scale=0.4]{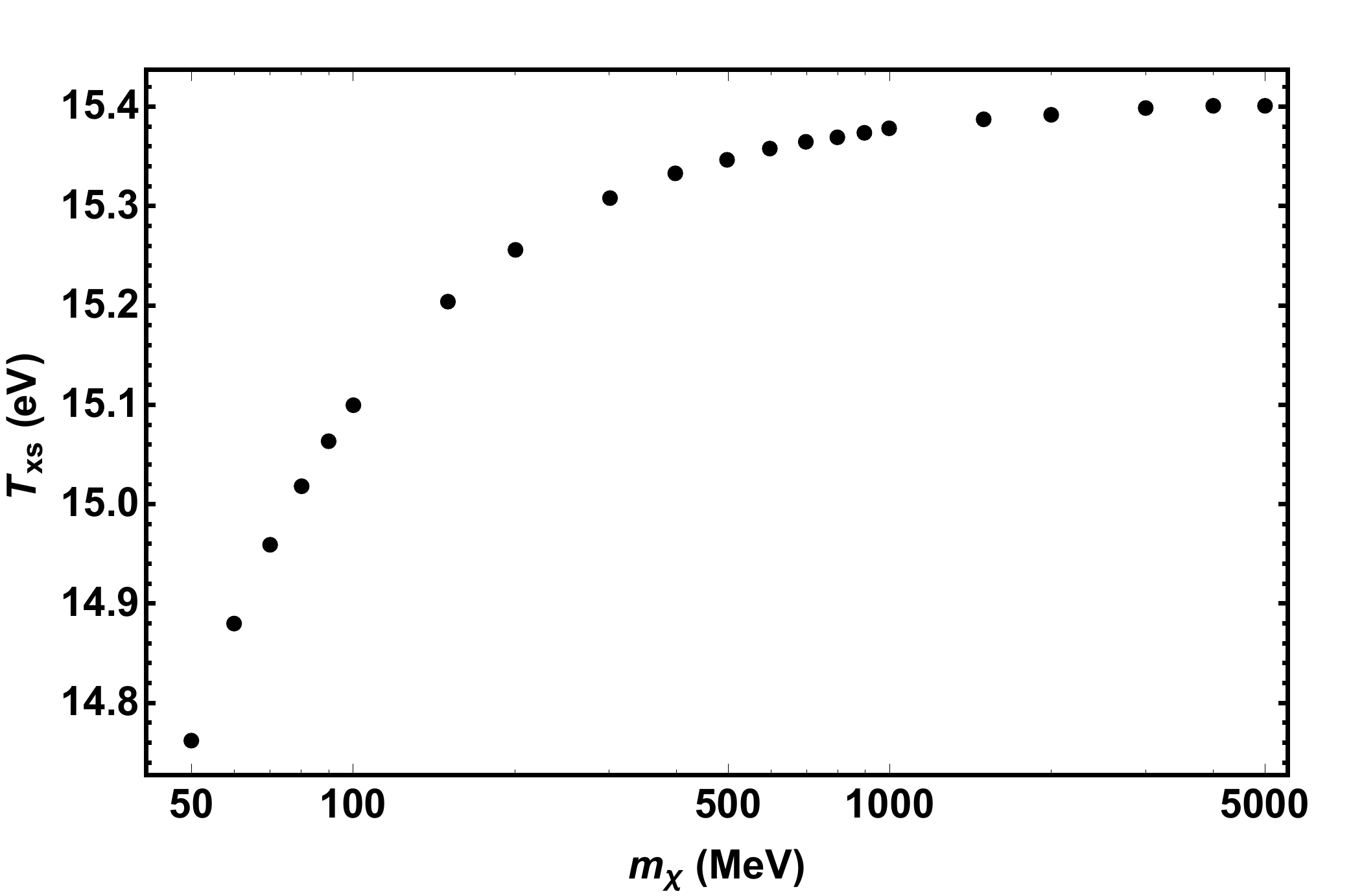} & \includegraphics[scale=0.4]{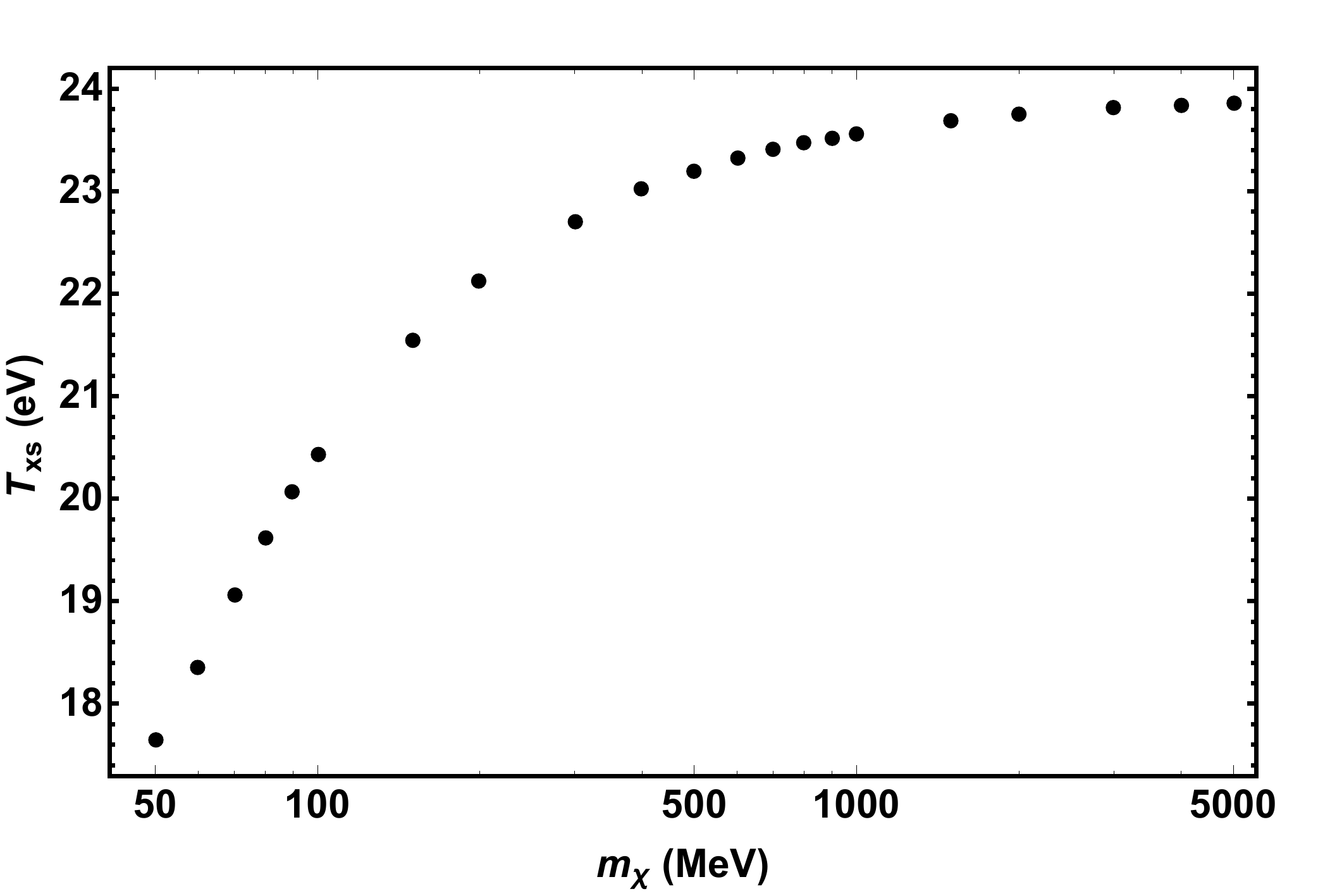}\tabularnewline
\end{tabular} 

\protect\caption{Energy transfer of DM, $T_{\textrm{xs}}$, below which scattering
with electron yield bigger $d\sigma/dT$ than proton in the hydrogen
ionization (assuming the same $\chi e$ and $\chi p$ coupling strengths),
plotted against DM mass $m_{\chi}$; (Left) for the $c_{1}$-type
and (Right) for the $d_{1}$-type interaction terms. \label{fig:c1_cross} }
\end{figure}

As DM interactions with electrons and nucleons are both included in
our calculations, it is interesting to compare their contributions.
Assuming the same coupling constants, $c_{1}^{(e)}=c_{1}^{(p)}=1/\textrm{MeV}^{2}$,
$d_{1}^{(e)}=d_{1}^{(p)}=1$, the comparison shown in Fig.~\ref{fig:c1_comp}
gives several important features:
\begin{enumerate}
\item In elastic scattering, the nuclear contribution dominates, and is
bigger than the electronic part by several orders of magnitude. Therefore,
elastic scattering is not likely to be a good channel of constraining
the LO DM-electron interactions, if the LO DM-nucleon interactions
are present and not unnaturally suppressed. 
\item In discrete excitations, the nuclear and electronic contributions
have sharp crossovers at energies slightly bigger than excitation
energies. If a detector is able to resolve these peaks where electronic
contributions clearly dominate, then it can be useful for setting
more stringent limits on the LO DM-electron interactions.
\item In ionization processes, unlike discrete excitations, the electronic
contributions generally dominate over the nuclear parts up to some
$T$ beyond the ionization thresholds. As a result, the LO DM-electron
interactions can hopefully be constrained in broader kinematic regions. 
\end{enumerate}
In Fig.~\ref{fig:c1_cross}, we study the $m_{\chi}$-dependence
of the crossover energy below which the DM-electron cross section
begins to be bigger than the DM-proton one (assuming the same coupling
strength) via the $c_{1}$- or $d_{1}$-type interaction that gives
rise to hydrogen ionization. The first thing to notice is in both
types of interactions and the considered range of $m_{\chi}$ (50
MeV to 5 GeV), there exist certain ranges of DM energy transfer $T$
where the ionization processes are more sensitive to the DM-electron
interaction than the DM-nucleon one. Furthermore, one observes that
the crossover energy for the $d_{1}$-type interaction is larger than
the one for the $c_{1}$-type interaction. The main reason is the
$1/q^{4}$ factor appearing in the double differential cross section
gives more weight to the response function at low $q^{2}$, which
enhances the role of electrons on one hand and suppresses the role
of nucleons on the other. 

Therefore, it is reasonable to conclude that the best observational
window to look for the LO DM-electron interactions is ionization processes
near threshold, in particular for LDM with $m_{\chi}<M_{H}$. The
discrete excitation peaks (which need good energy resolution of detectors)
also provide good supplements. Although hydrogen can hardly be a good
candidate for detecting LO DM-electron interactions for the low energy
transfer $T\sim10-20\,\textrm{eV}$ is far below the current detector
thresholds, however, our above conclusion makes good sense for practical
detector species made of heavy atoms: Not only the ionization thresholds
of atomic inner orbitals can be as high as a few or tens of keV which
are observable in current detectors, but also there are more than
one ionization peaks which can provide additional information.

\subsection{NLO Interactions of $c_{11}$, $d_{11}$, $c_{10}$, and $d_{10}$}

\begin{figure}[h]
\begin{tabular}{cc}
\includegraphics[scale=0.4]{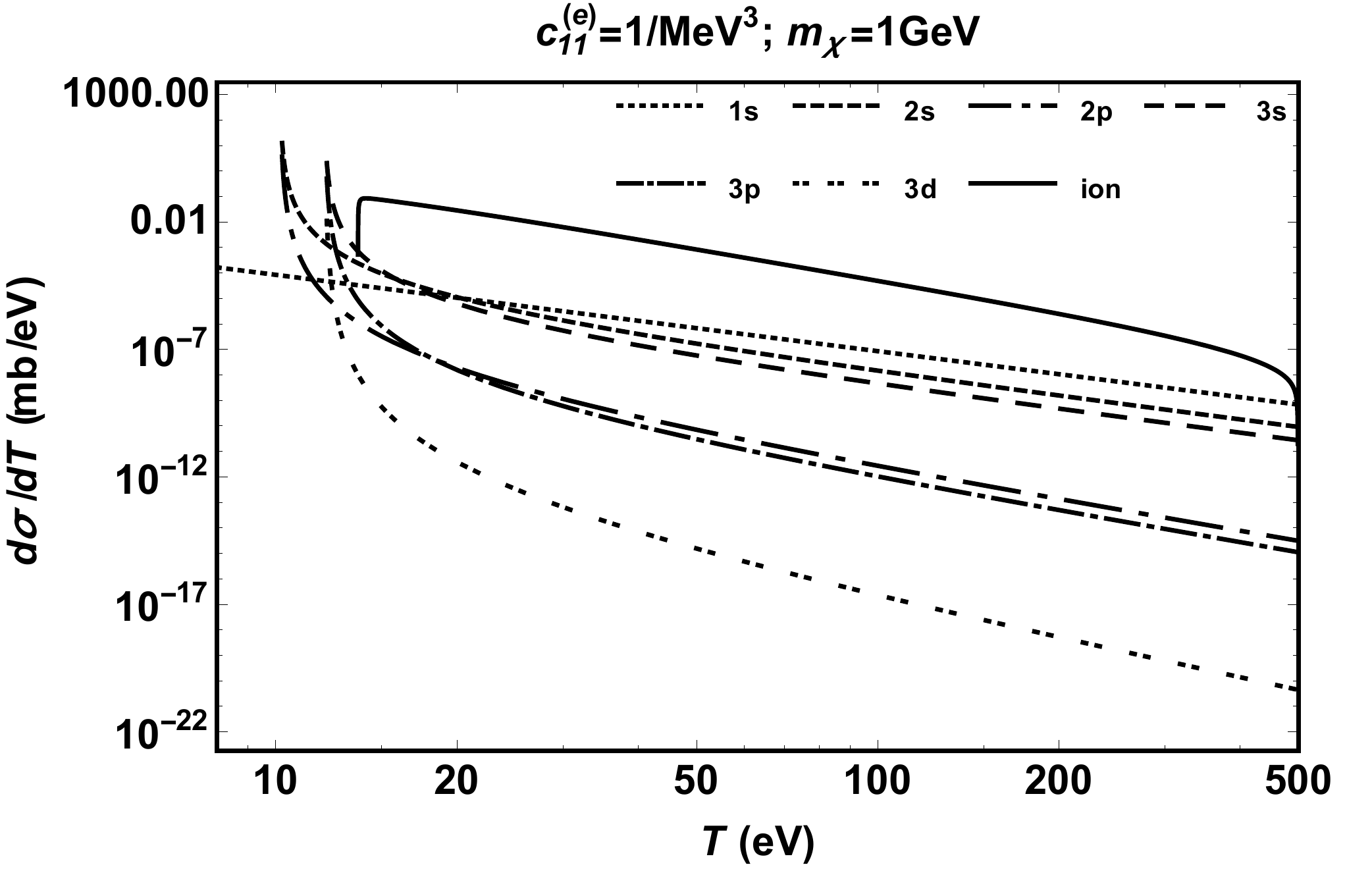} & \includegraphics[scale=0.4]{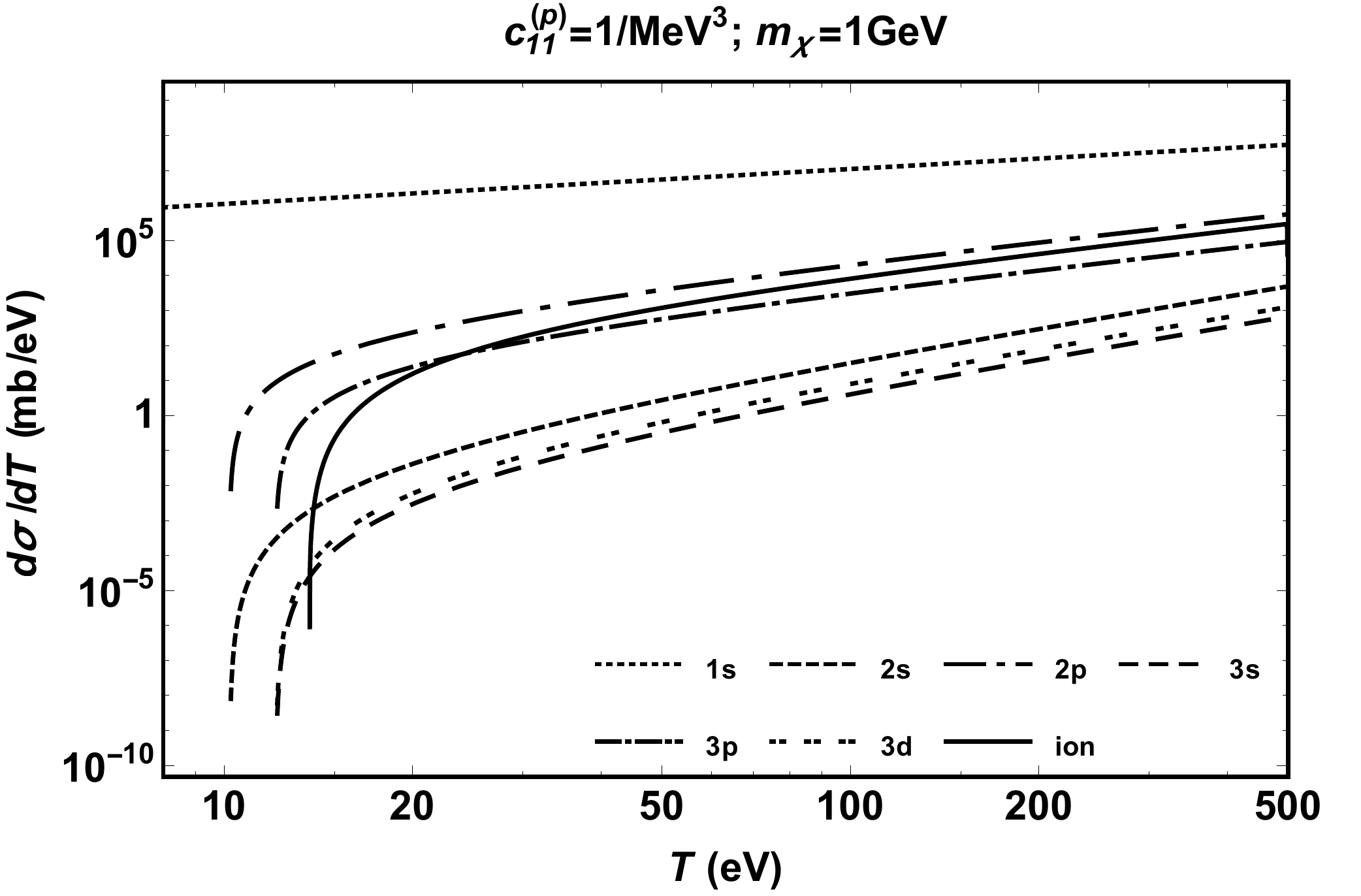}\tabularnewline
\includegraphics[scale=0.4]{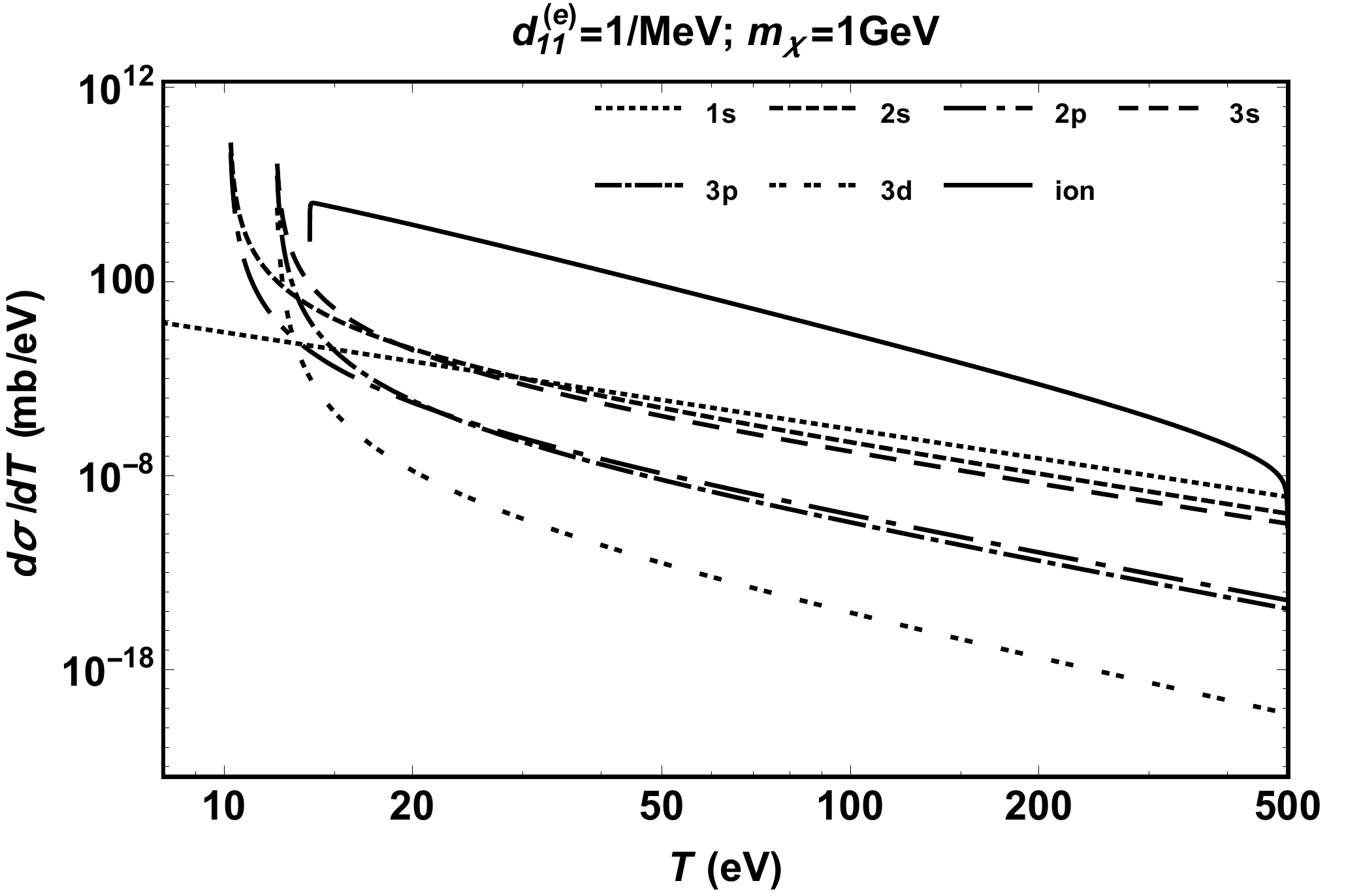} & \includegraphics[scale=0.4]{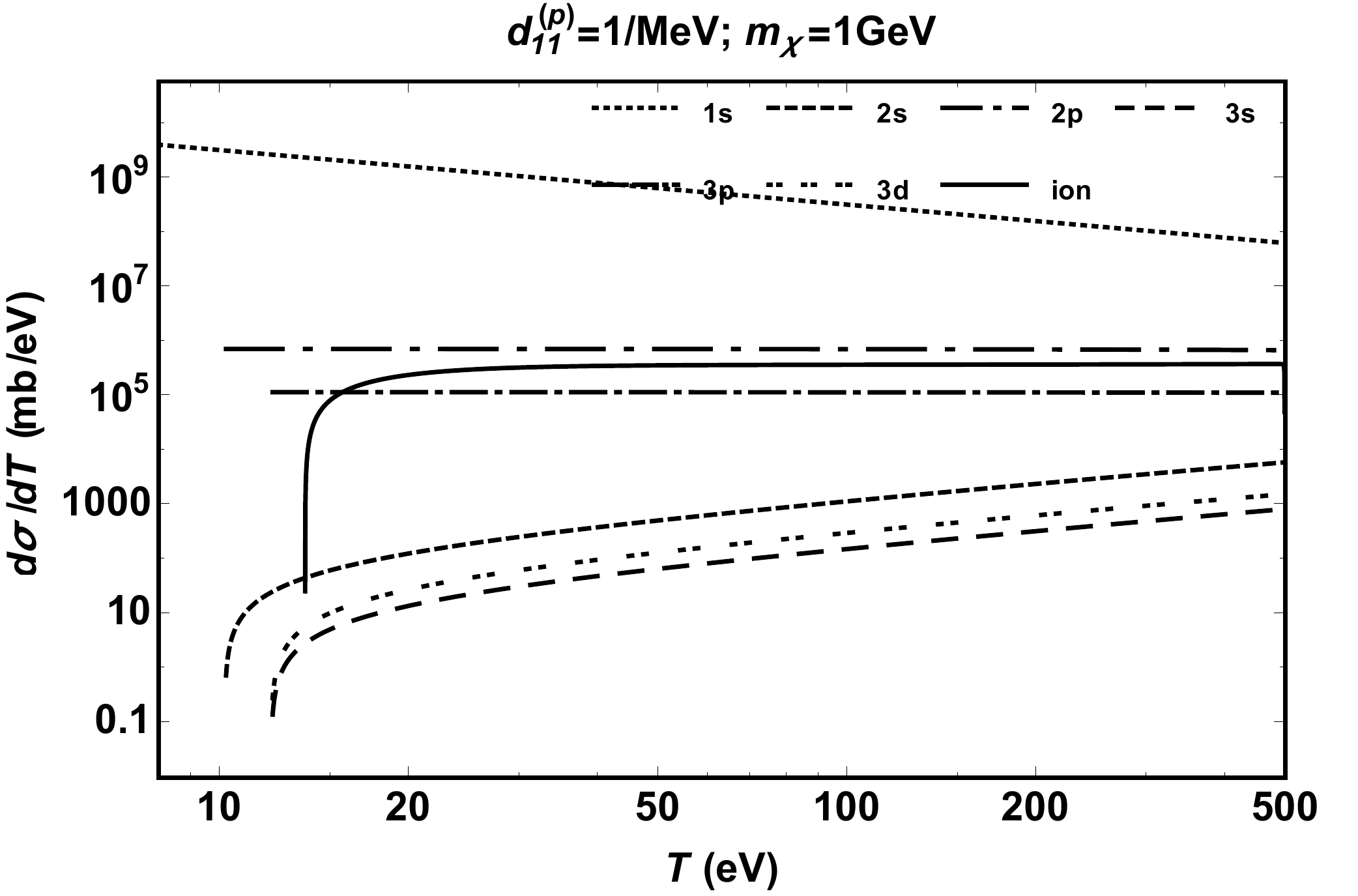}\tabularnewline
\end{tabular} 

\protect\caption{Different channels of DM scattering with $m_{\chi}=1\,\mathrm{GeV}$
and interaction of $c_{11}^{(e)}=1/\textrm{MeV}^{3}$ (upper-left),
$c_{11}^{(p)}=1/\textrm{MeV}^{3}$ (upper-right), $d_{11}^{(e)}=1/\textrm{MeV}$
(lower-left), $d_{11}^{(p)}=1/\textrm{MeV}$ (lower-right). \label{fig:c11_1GeV} }
\end{figure}

Consider now the interaction terms of $c_{11}$ and $d_{11}$, the
results for $m_{\chi}=1\,\mathrm{GeV}$ are presented in Figs.~\ref{fig:c11_1GeV}.
The main change from the corresponding plots of $c_{1}$ and $d_{1}$
is the extra $q^{2}$ factor appearing in the differential cross sections.
For elastic scattering or discrete excitations away from threshold,
this factor introduces an extra dependence on $T$, and again can
be best seen by a comparison of the $c_{1}^{(p)}$ and $c_{11}^{(p)}$
curves for elastic scattering. For ionization, the impact of the $q^{2}$
factor, which is to be integrated over some allowed range, however
can not be easily factored out.

Our previous argument that $d\sigma/dT$'s of elastic scattering and
discrete excitations are independent of $m_{\chi}$ also applies to
the cases of $c_{11}$ and $d_{11}$ (also $c_{10}$ and $d_{10}$
to be discussed later). Therefore, we do not repeat the results for
$m_{\chi}=50\,\textrm{MeV}$ and just point out that they are the
same as for the $m_{\chi}=1\,\textrm{GeV}$ case in the regions bounded
by the cutoff energies given by Eq.~(\ref{eq:T_cut}). Also, while
the ionization processes do have $m_{\chi}$ dependence, it does not
differ from what has been shown in Fig.~\ref{fig:c1_50MeV} for the
case of $c_{1}$ and $d_{1}$ in a significant way.

Similarly to the $c_{1}$- and $d_{1}$-type interactions with DM,
elastic scattering is the best to constrain the $c_{11}^{(p)}$ and
$d_{11}^{(p)}$ terms, while inelastic scattering at discrete excitation
peaks and of ionization are more suitable for the $c_{11}^{(e)}$
and $d_{11}^{(e)}$ terms. To further disentangle the dependence of
$d\sigma/dT$ on $c_{1}$, $d_{1}$, $c_{11}$, and $d_{11}$, the
scaling of $d\sigma/dT$ with energy transfer $T$ can provide useful
guidance: For example, in elastic scattering and discrete excitations
away from thresholds, the energy dependence of $d\sigma/dT$ on the
$|c_{1}|^{2}$, $|d_{1}|^{2}$, $|c_{11}|^{2}$, and $|d_{11}|^{2}$
terms is $T^{0}$, $T^{-2}$, $T^{1}$, and $T^{-1}$, respectively. 

\begin{figure}[h]
\begin{tabular}{cc}
\includegraphics[scale=0.4]{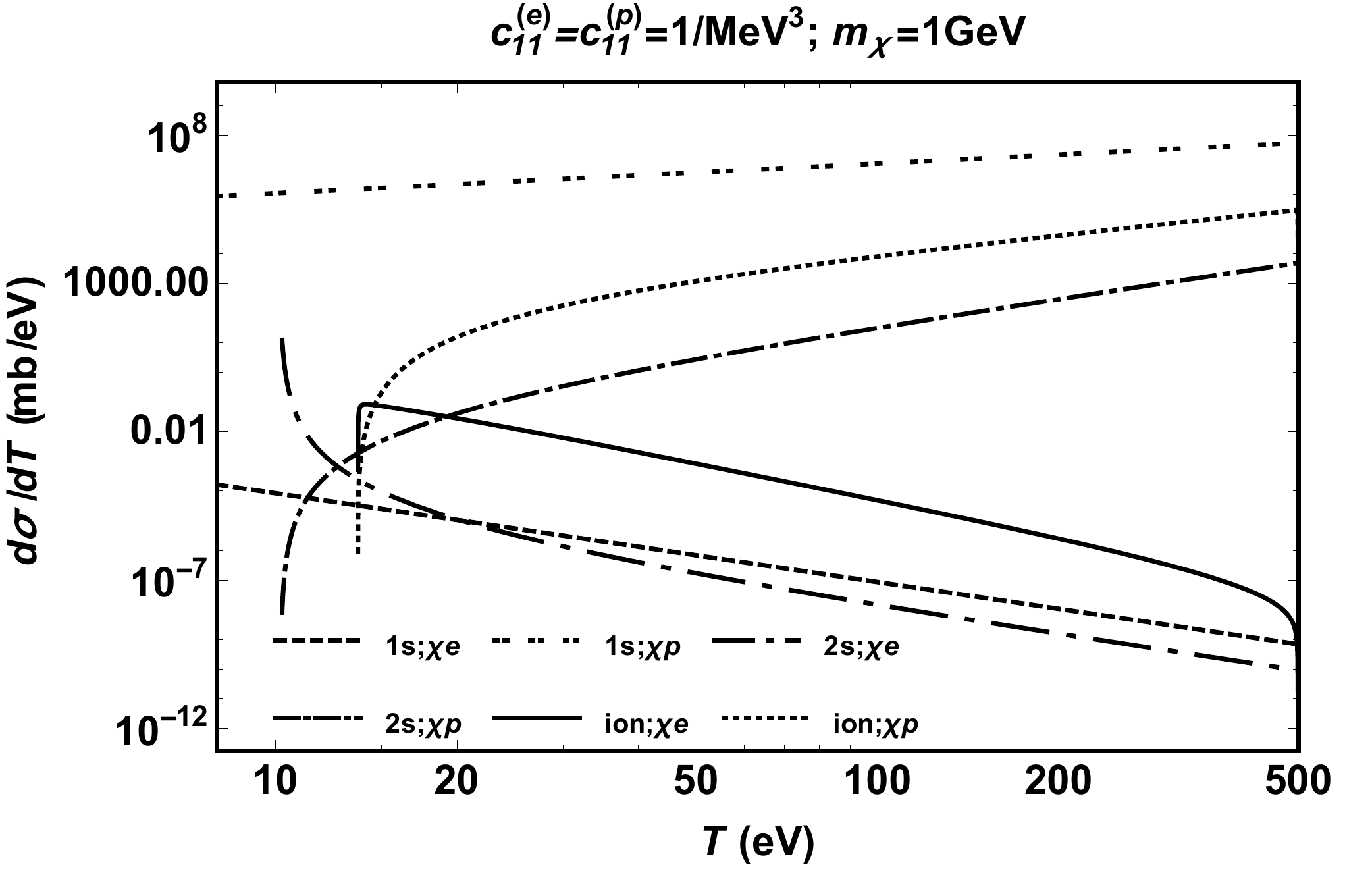} & \includegraphics[scale=0.4]{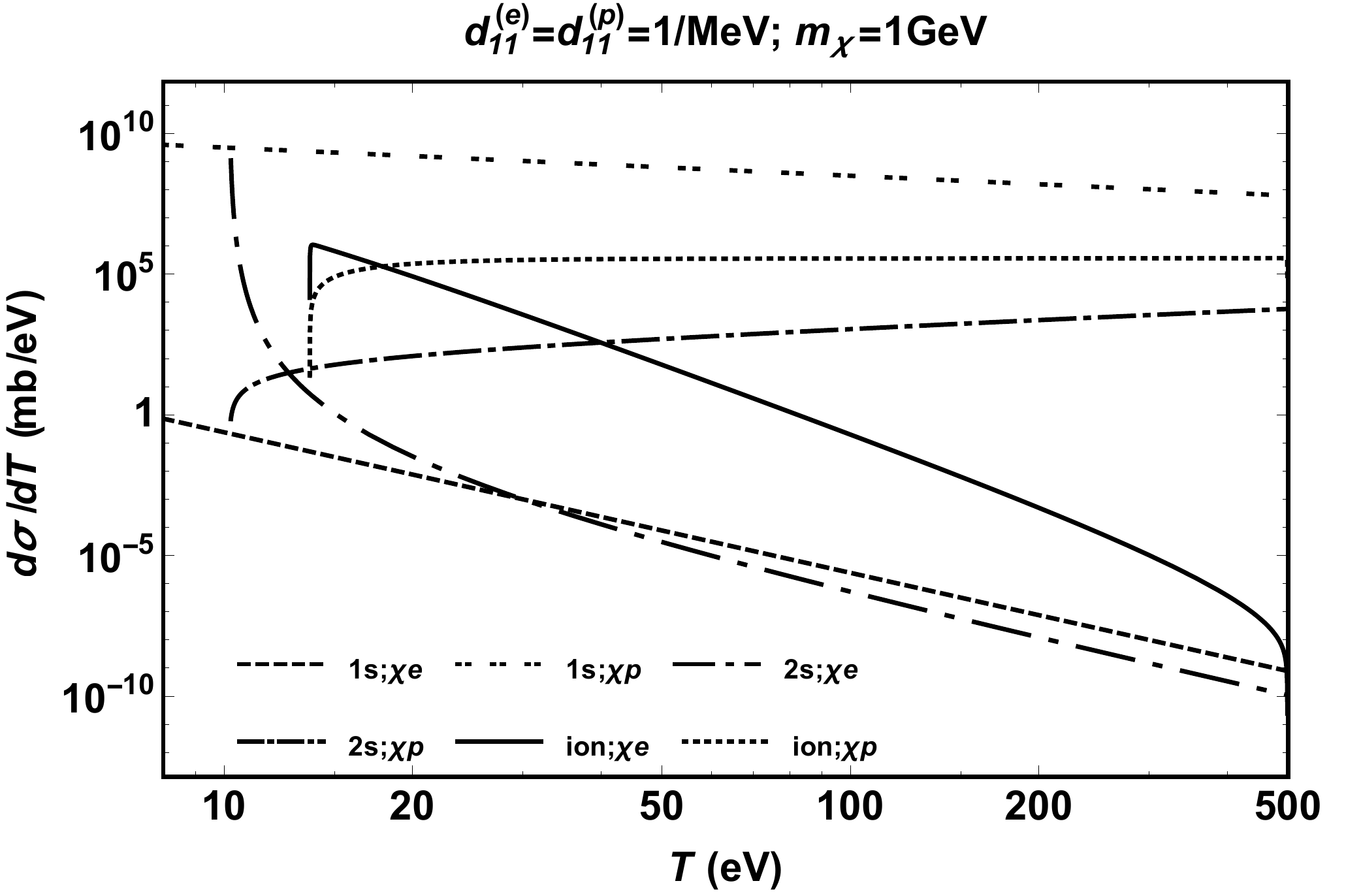}\tabularnewline
\end{tabular} 

\protect\caption{Comparison of DM cross sections with the electron ($\chi e$) and
proton ($\chi p$) in a hydrogen atom for selected channels including
(i) elastic (1s), (ii) discrete excitation to 2s, and (iii) ionization
(ion). The interactions are taken to be (Left) $c_{11}^{(e)}=c_{11}^{(p)}=1/\textrm{MeV}^{3}$
and (Right) $d_{11}^{(e)}=d_{11}^{(p)}=1/\textrm{MeV}$. Interference
terms due to $\chi e$ and $\chi p$ amplitudes are ignored. \label{fig:c11_comp} }
\end{figure}

The pattern regarding the competition of electronic and nuclear contributions
in discrete excitations and ionization with the $c_{11}$ and $d_{11}$
terms is similar to the case with the $c_{1}$ and $d_{1}$ terms:
sharp crossovers in discrete excitations and some ranges of electronic
dominance near ionization threshold. Some examples are given in Fig.~\ref{fig:c11_comp}. 

\begin{figure}[h]
\begin{tabular}{cc}
\includegraphics[scale=0.4]{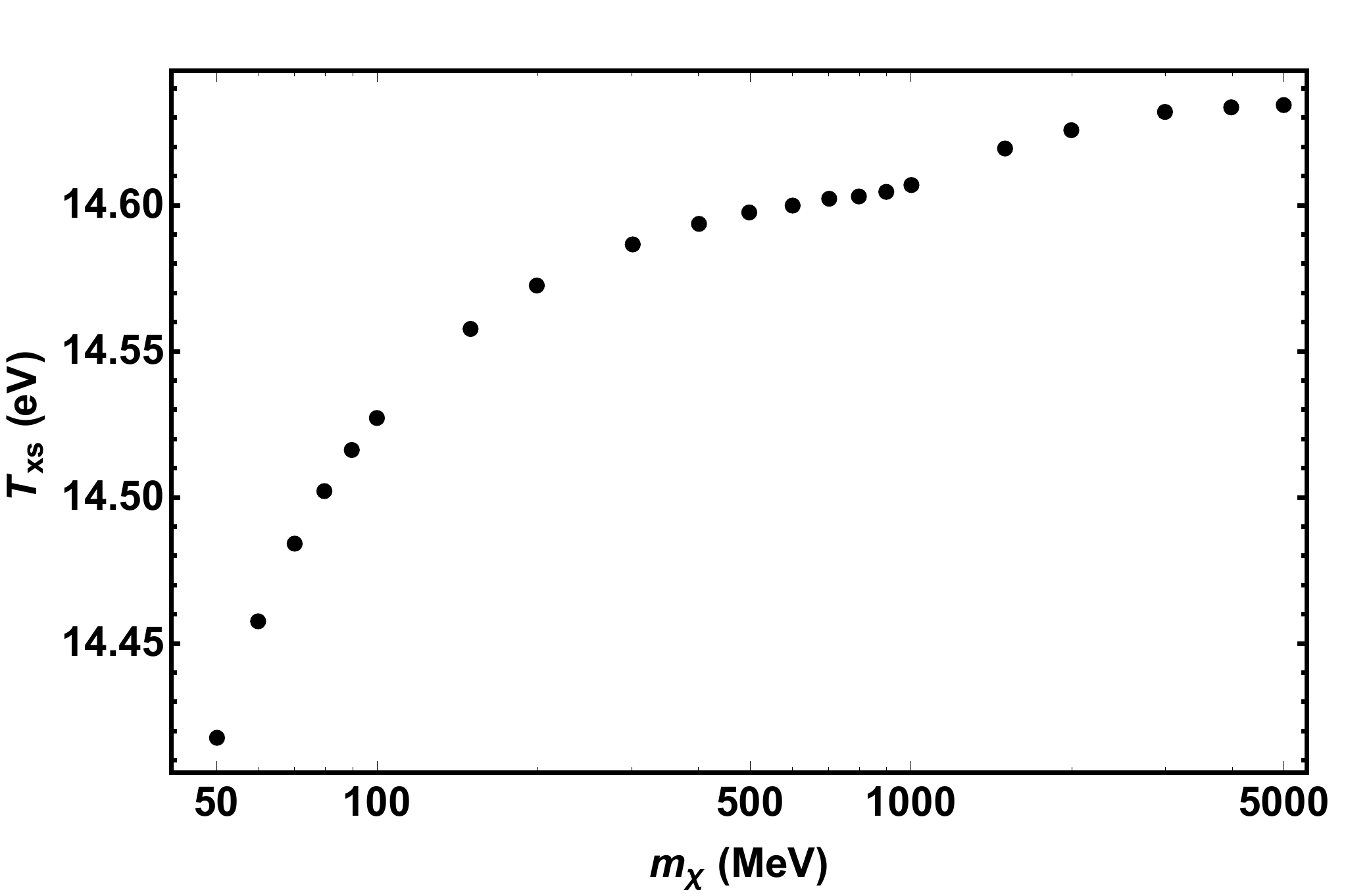} & \includegraphics[scale=0.4]{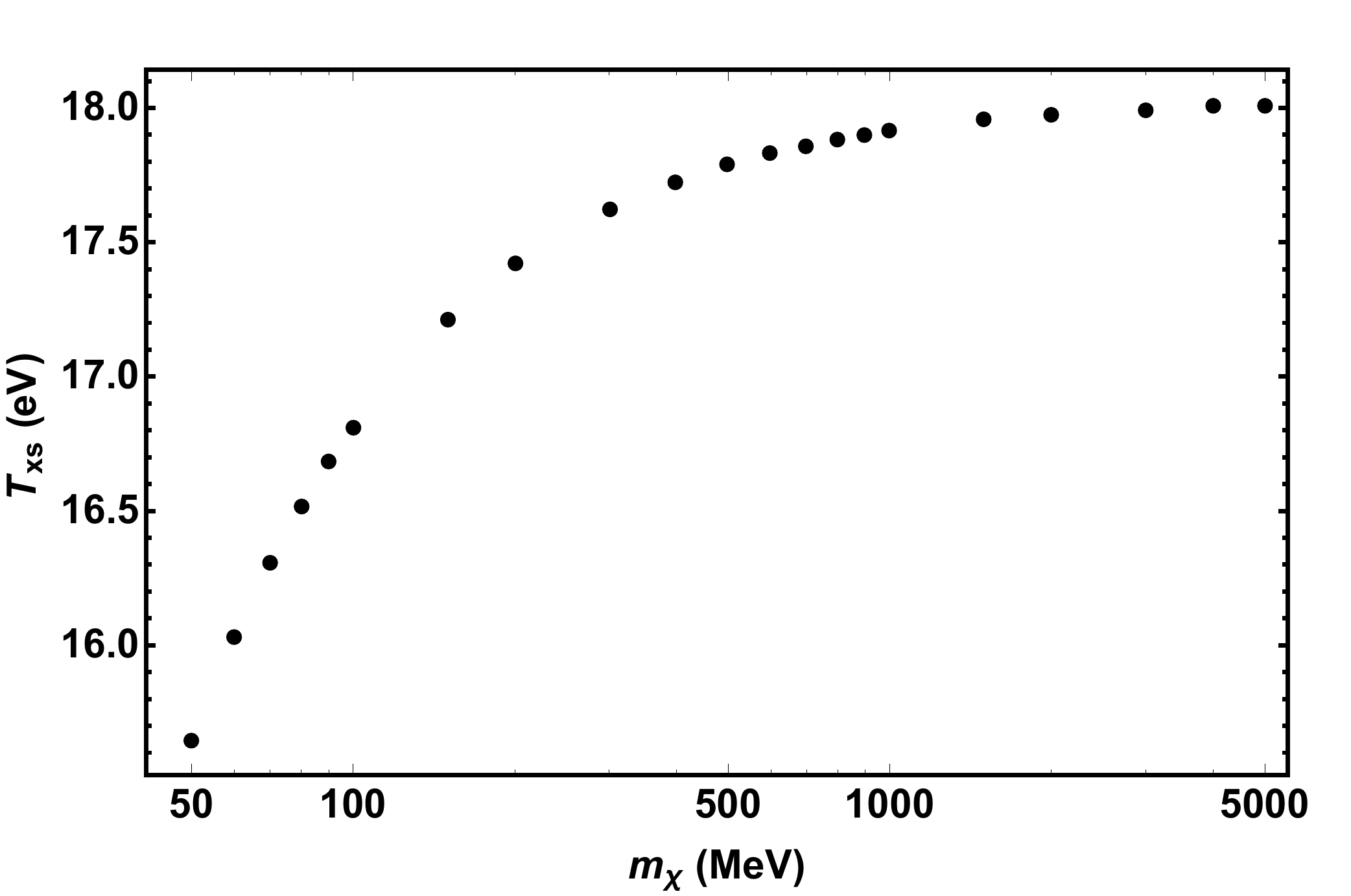}\tabularnewline
\end{tabular} 

\protect\caption{Energy transfer of DM, $T_{\textrm{xs}}$, below which scattering
with electron yield bigger $d\sigma/dT$ than proton in the hydrogen
ionization (assuming the same $\chi e$ and $\chi p$ coupling strengths),
plotted against DM mass $m_{\chi}$; (Left) for the $c_{11}$-type
and (Right) for the $d_{11}$-type interaction terms. \label{fig:c11_cross} }
\end{figure}

The energy transfer $T_{\textrm{xs}}$ below which the electronic
contribution is bigger than the nuclear one, assuming $c_{11}^{(e)}=c_{11}^{(p)}=1/\textrm{MeV}^{3}$
and $d_{11}^{(e)}=d_{11}^{(p)}=1/\textrm{MeV}$, in the ionization
processes is plotted in Fig.~\ref{fig:c11_cross} against $m_{\chi}$.
Notice that the values of $T_{\textrm{xs}}$ for the $c_{11}$- and
$d_{11}$-type interactions at a given $m_{\chi}$ are both reduced
in comparison with the cases of the LO $c_{1}$ and $d_{1}$ terms,
respectively. This is in agreement with the expectation that the extra
$q^{2}$ factor in the double differential cross section reduces the
weight of the small $q^{2}$ region, so the electronic contribution
is relatively suppressed than the nuclear part. 

Similar to the LO case, the contributions from the target-spin-dependent
$c_{10}$ and $d_{10}$ terms can be obtained from the target-spin-independent
results of $c_{11}$ and $d_{11}$, simply by adding factors due to
spin matrix elements (see Eq.~\ref{eq:spin_mat}). Therefore, all
observations and conclusions made in the $c_{11}$ and $d_{11}$ cases
apply to the $c_{10}$ and $d_{10}$.

However, regarding the competition between the electronic and nuclear
contributions in scattering processes involving the $c_{10}$ or $d_{10}$
term, there is a subtlety arising from the natural scales of $c_{10}^{(e)}/c_{10}^{(p)}$
and $d_{10}^{(e)}/d_{10}^{(p)}$. If one takes the point of view that
both EFT interaction terms of electrons and nucleons are matched to
a more fundamental theory at some high scale $\Lambda$, then it is
reasonable to anticipate the possibility that $c_{10}^{(e)}/c_{10}^{(p)}\sim1$
and $d_{10}^{(e)}/d_{10}^{(p)}\sim1$. On the other hand, the masses
of an electron and a nucleon differ by three orders of magnitude.
If the $c_{10}$ and $d_{10}$ terms are matched to a relativistic
theory, for example, $(\bar{\chi}\chi)(\bar{f}i\gamma_{5}f)$ and
$(\bar{\chi}\chi)(\bar{f}i\gamma_{5}f)/q_{\mu}^{2}$ at some high
scale, the resulting nonrelativistic EFT expansion at NLO will involve
the expansion parameter $q/m_{f}$ to first order. In such cases,
then one should expect $c_{10}^{(e)}/c_{10}^{(p)},\,d_{10}^{(e)}/d_{10}^{(p)}\sim m_{p}/m_{e}\sim2\times10^{3}$.
This in turn would largely increase the sensitivity of discrete excitation
peaks and ionization processes on the NLO DM-electron interaction
terms such as $c_{10}$ and $d_{10}$. 

\begin{figure}[h]
\begin{tabular}{c}
\includegraphics[scale=0.4]{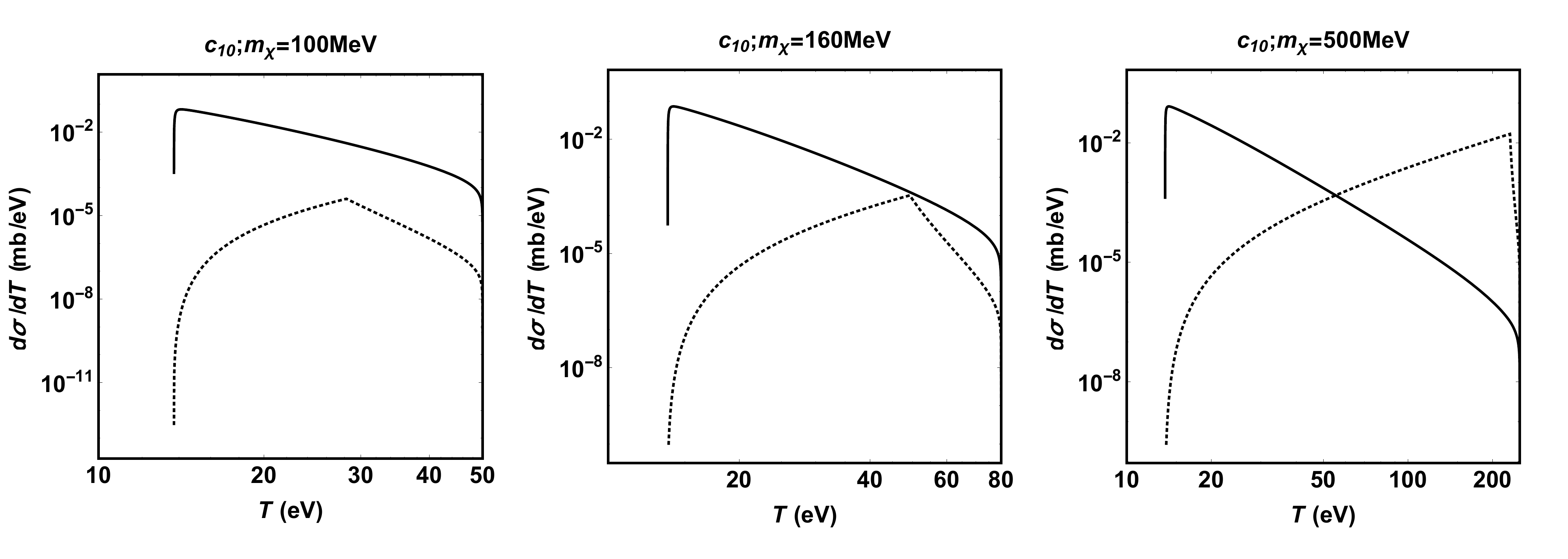}\tabularnewline
\includegraphics[scale=0.4]{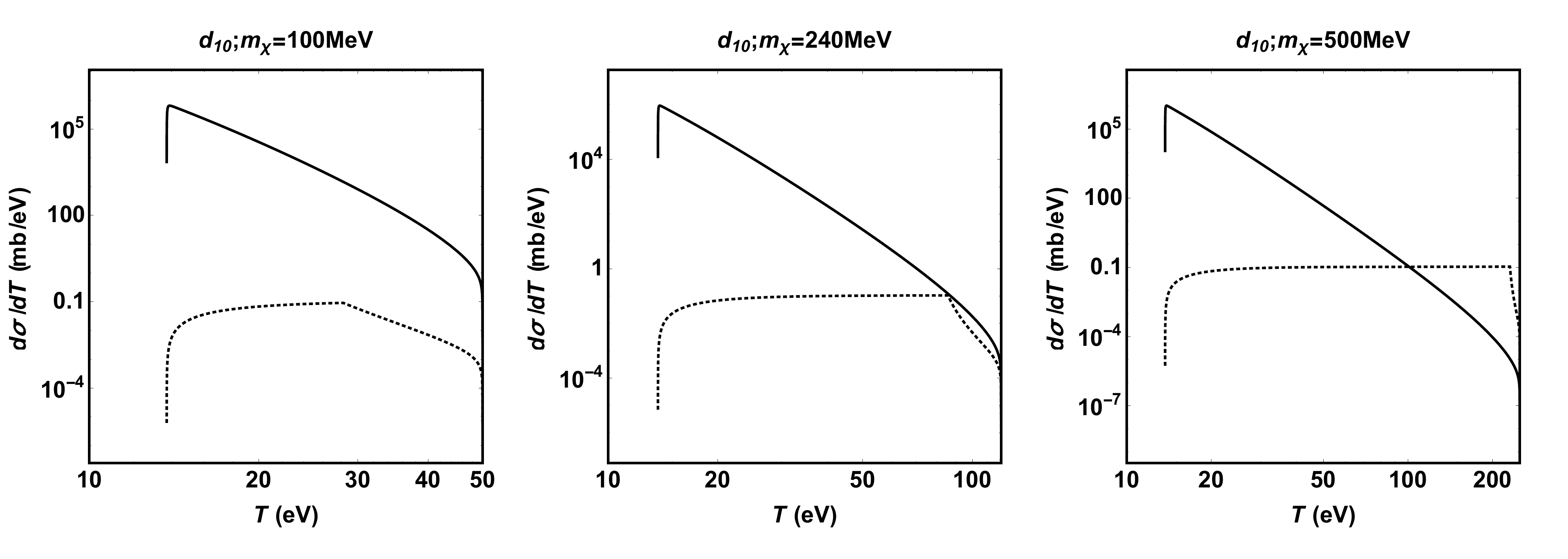}\tabularnewline
\end{tabular} 

\protect\caption{Comparison of electronic (solid lines) and nuclear (dotted lines)
contributions to $d\sigma/dT$ for selected DM mass $m_{\chi}$ with
(Upper) $c_{10}^{(e)}=c_{10}^{(p)}m_{p}^{2}/m_{e}^{2}=1/\textrm{MeV}^{3}$
and (Lower) $d_{10}^{(e)}=d_{10}^{(p)}m_{p}^{2}/m_{e}^{2}=1/\textrm{MeV}$.
\label{fig:c10_cross} }
\end{figure}

An example is given in Fig.~\ref{fig:c10_cross}: For $m_{\chi}\lesssim160\,\text{MeV}$
and $m_{\chi}\lesssim240\,\text{MeV}$ respectively for the $c_{10}$
and $d_{10}$ terms, the electronic contributions are larger than
the nuclear ones in the entire allowed ranges of $T\le\nicefrac{1}{2}m_{\chi}v_{\chi}^{2}$.
For heavier $m_{\chi}$, the crossovers both happen at energies further
away from ionization thresholds, $\sim50\,\textrm{eV}$ and $100\,\textrm{eV}$
respectively for the $c_{10}$ and $d_{10}$ terms -- much bigger
than other interactions terms previously discussed. 

Briefly concluding this subsection, we point out that the best observational
window to look for the NLO DM-electron interactions terms including
$c_{11}$, $d_{11}$, $c_{10}$, and $d_{10}$ is still the ionization
processes near threshold and the discrete excitation peaks. The different
energy dependence of $d\sigma/dT$ from the LO terms in principle
provides a way to disentangle them. Furthermore, because of the huge
difference between the masses of an electron and a nucleon, interaction
terms that depend on the relativity of scattered particles can be
further separated. In most situations, such NLO DM-electron interactions
can be sensitively constrained without much background arising from
similar DM-nucleon interactions because atomic electron can be very
relativistic while atomic nuclei and nucleons inside are mostly nonrelativistic.

\section{Summary \label{sec:summary}}

In this paper, we study the scattering processes of sub-GeV DM particles
and hydrogen atoms, including elastic, atomic discrete excitation,
and atomic ionization channels. The interactions of DM with electrons
and nucleons are both included and formulated in a general framework
based on nonrelativistic effective field theory. In addition to the
leading-order spin-independent and spin-dependent contact terms, we
also include the possibility of long-ranged DM interactions and a
few next-to-leading-order terms. Some of the interaction terms yield
orthogonal scattering amplitudes, but there are also interference
terms. Disentanglement of various interaction terms can in principle
be done by their different dependence on DM energy deposition in scattering
cross sections.

On the assumption of same dark matter coupling strengths, it is found
that DM--electron interactions dominate the inelastic transitions
to discrete excited states and ionization continuum around their threshold
regions (sizes of these regions depend on interaction types), and
DM--nucleon interactions become more important with increasing energy
and dominate in elastic scattering. These conclusions can be used
to guide the searches of sub-GeV DM interactions in optimal experimental
configurations and kinematics. For DM--electron interactions, the
inelastic peaks of discrete excitations and ionizations in scattering
cross sections, which can be taken as smoking-gun signals of DM scattering,
can further increase an experiment's constraining power. For DM--nucleon
interactions, although the elastic scattering is the best channel,
however, for light DM particles which can not deposit observable energies
in detectors, one has to rely on the high energy part of ionization
processes. 

The energy and momentum transfers involved in sub-GeV DM scattering
overlap typical atomic scales, so studies of issues such as binding
effects and electron/nuclear recoil mechanism, which play important
roles in interpreting experimental data, require detailed many-body
calculations. This case study of hydrogen, where both binding and
recoil can be taken into account most simply, therefore provides useful
qualitative understanding of what to be anticipated in sub-GeV DM
scattering off practical detector materials such as germanium and
xenon.
\begin{acknowledgments}
We acknowledge the support from the Ministry of Science and Technology
of Republic of China under Grants No. 102-2112-M-002-013-MY3 (J.-W.
C., C.-L. W., and C.-P. W.) and No. 103-2112-M-259-003 (H.-C. C. and
C.-P. L.); the Center for Theoretical Sciences and Center of Advanced
Study in Theoretical Sciences of National Taiwan University (J.-W.
C., C.-L. W., and C.-P. W.); and the National Center for Theoretical
Sciences. J.-W. C. was also supported in part by the Deutsche Forschungsgemeinschaft
and National Natural Science Foundation of China (CRC 110).
\end{acknowledgments}

\bibliography{draft}

\end{document}